\documentclass[%preprint, %twocolumn,
prd,amsmath,aps,floats,epsf,epsfig,nofootinbib,letter]{revtex4-1}

\usepackage{hyperref}
\usepackage{graphicx}
\usepackage{subfigure}

\usepackage{bm}
\usepackage{array}
\usepackage{dcolumn}
\usepackage{verbatim}
\makeatletter

\newcommand{\Rmnum}[1]{\expandafter\@slowromancap\romannumeral #1@}
\makeatother

\newcommand{\be}{\begin{eqnarray}}
\newcommand{\en}{\end{eqnarray}}
\newcommand{\non}{\nonumber}
\newcommand{\DGammas}{\Delta \Gamma_s}
\newcommand{\DGamma}{\Delta \Gamma}
\newcommand{\Bbar}{\bar{B}}
\newcommand{\Dbar}{\bar{D}}
\newcommand{\Kbar}{\bar{K}}
\newcommand{\Dst}{D^{*}}
\newcommand{\Kst}{K^{*}}

\newcommand{\Tr}{\mathcal T}
\newcommand{\Cur}{\mathcal J}
\newcommand{\CP}{\textit{CP}}
\newcommand{\Br}{\mathcal{B}}
\newcommand{\eps}{\varepsilon}
\newcommand{\DDKgeneral}{D^{(*)}_s\Dbar^{(*)}\Kbar^{(*)}}

\newcommand{\err}[2]{$#1 \pm #2$}
\newcommand{\errr}[3]{$#1 \pm #2 \pm #3$}
\newcommand{\NTU}{Department of Physics, National Taiwan University,
    Taipei 10617, Taiwan (R.O.C.)}
\newcommand{\NCTSnorth}{
National Center for Theoretical Sciences, National Taiwan University,
Taipei, Taiwan 10617 (R.O.C.)}
\newcommand{\CYCU}{Department of Physics, Chung Yuan Christian University,
    Chung-Li 32023, Taiwan (R.O.C.)}

\begin{document}

\title{
Long-Distance Contribution to $\DGamma_s$ of the $B_{s}-\Bbar_{s}$ System}

\author{Chun-Khiang Chua}
\affiliation{\CYCU}
\author{Wei-Shu Hou}
\affiliation{\NTU,\NCTSnorth}
\author{Chia-Hsien Shen}\affiliation{\NTU,\NCTSnorth}

\date{\today}

\begin{abstract}
We estimate the long-distance contribution to
the width difference of $B_s-\Bbar_s$ system, based mainly on
two-body $D^{(*)}_s\Dbar^{(*)}_s$ modes and
three-body $\DDKgeneral$ modes (and their $CP$ conjugates).
Some higher $c\bar{s}$ resonances are also considered.
The contribution to $\DGamma_s/ \Gamma_s$ by two-body modes
is $(10.2\pm3.0)\%$,
slightly smaller than the short-distance result of $(13.3 \pm 3.2)\%$.
The contribution to $\DGamma_s/ \Gamma_s$ by $D_{s0}^*(2317)$, $D_{s1}(2460)$,
and $D_{s1}(2536)$ resonances is negligible.
For the three-body $D^{(*)}_s\Dbar^{(*)}\Kbar^{(*)}$ modes,
we adopt the factorization formalism and model the form factors
with off-shell $D^{(*)}_s$ poles, the $D_{sJ}(2700)$ resonance,
and non-resonant (NR) contributions.
These three-body modes can arise through current-produced or transition
diagrams, but only SU(3)-related $D^{(*)}_{u,d} \Dbar^{(*)}\Kbar$ modes
from current diagram have been measured so far.
The pole model results for $D^*_{u,d} \Dbar^{(*)}\Kbar$
agree well with data, while $D_{u,d} \Dbar^{(*)}\Kbar$ rates
agree with data only within a factor of 2 to 3.
All measured $D^{(*)}_{u,d} \Dbar^{(*)}\Kbar$ rates can be reproduced
by including NR contribution.
The total $\DGamma_s/\Gamma_s$ obtained is $(16.7 \pm 8.5) \%$,
which agrees with the short-distance result within uncertainties.
For illustration, we also demonstrate the effect of $D_{sJ}(2700)$ in modes
with $D^{(*)}\Kst$.
In all scenarios, the total $\DGamma_s/\Gamma_s$ remain consistence
to the short-distance result.
Our result indicates that
(a)~the operator product expansion (OPE) in short-distance picture
is a valid assumption,
(b)~approximating
the $B_s \to D^{(*)}_s\Dbar^{(*)}_s$ decays to saturate $\DGamma_s$
has a large correction,
(c)~the effect of three-body modes cannot be neglected,
and
(d)~in addition to $D_s$ and $D^*_s$ poles, the $D_{sJ}(2700)$ resonance
 also plays an important role in three-body modes.
Future experiments are necessary to improve the estimation of $\DGamma_s$
from long-distance picture.
\end{abstract}

\maketitle

\section{Introduction and Motivation}
\label{sec:Intro}

One of the most exciting news in particle physics last year is the
anomalous like-sign dimuon charge asymmetry $A^b_{sl}$ reported by
the D0 collaboration~\cite{Abazov:2010hv}.
The updated result is
$A^b_{sl} =(-0.787 \pm0.172\; \text{(stat)} \pm0.093\; \text{(syst)})\%$,
based on $9.0\,\text{fb}^{-1}$ data~\cite{Abazov:2011}.
The result is $3.9 \sigma$ larger than the Standard Model (SM) prediction
of $(-0.024\pm0.003)\%$~\cite{Lenz:2007JHEP}.
This asymmetry is comprised by the wrong-sign asymmetries $a^{d,s}_{sl}$
for $B_{d,s}$ mesons~\cite{Abazov:2011, HFAG:2010},
%% Asl relation
\be
A^b_{sl} = (0.594 \pm 0.022)a^{d}_{sl}+(0.406 \pm 0.022)a^{s}_{sl}.
 \label{eq:Asl_relation}
\en
From direct measurements by B factories~\cite{HFAG:2010},
$a^{d}_{sl}=-(0.05 \pm 0.56) \%$
does not deviate from the SM prediction~\cite{Lenz:2007JHEP}.
Imposing these two experimental values into Eq.~(\ref{eq:Asl_relation}),
one finds a large $a^{s}_{sl}$.
The very recent update used muon impact parameter
to directly extract~\cite{Abazov:2011}
\begin{equation}
a^{d}_{sl}=-(0.12 \pm 0.52) \%,
\quad
a^{s}_{sl}=-(1.81 \pm 1.06) \%.
\end{equation}
The result of $a^{s}_{sl}$ is much larger than the SM prediction
of $(1.9 \pm 0.3)\times 10^{-5}$~\cite{Lenz:2007JHEP}.
The current world average of $a^{s}_{sl}$, done
before the very recent update~\cite{Abazov:2011},
is~\cite{HFAG:2010}
%% asl value
\begin{equation}
a^{s}_{sl} = -0.0115 \pm 0.0061,
\label{eq:asl_result}
\end{equation}
which is still much larger than the SM prediction.
This anomalous result has drawn intense theoretical attention, including
model-independent analyses~\cite{Ligeti:2010ia, Buras:2010mh,
Deshpande:2010hy, Bauer:2010dga, Chen:2010aq},
and explanations
from specific new physics models~\cite{Lenz:2010gu, Dobrescu:2010rh,
Chen:2010wv, Dighe:2010nj, SUSY:2010, Bai:2010kf, Dutta:2010ji, Oh:2010vc}.

The wrong-sign asymmetry $a^{s}_{sl}$ can be derived from
mixing parameters~\cite{Abazov:2010hv}
%% asl expression
\begin{equation}
a^s_{sl} = \frac{\Delta \Gamma_s}{\Delta m_s} \text{tan} \phi_s
         = \frac{2 \lvert \Gamma_{12,s} \rvert}{\Delta m_s} \text{sin} \phi_s,
  \label{eq:asl_relation}
\end{equation}
where the $\DGammas$ and $\Delta m_s$ are the width difference and mass
difference of $B_s-\Bbar_s$ system,
$\phi_s$ is the $CP$ violating phase,
and $\Gamma_{12,s}$
is the absorptive off-diagonal element of mixing matrix
(see Section~II.~A for more detail).
Note that $a^s_{sl}$ is bounded by $2 \lvert \Gamma_{12,s} \rvert/\Delta m_s$.
The short-distance calculation in SM predicts~\cite{Lenz:2007JHEP},
%% short-distance \DGamma
\be
\DGamma_{s,\text{SM}}  &= (0.087\pm 0.021)\ \text{ps}^{-1},
\non\\
\DGamma_{s,\text{SM}} / \Gamma_{s,\text{SM}} & = (13.3 \pm 3.2) \%,
\non\\
\Delta m_{s,\text{SM}} &            = (17.3\pm 2.6)\ \text{ps}^{-1}, \label{eq:DGamma_sd}\\
\phi_{s,\text{SM}} &            = (0.22^{\circ}\pm0.06^{\circ}).
\non
\en
Note that $\phi_s$ is very small in SM,
so $2\lvert \Gamma_{12,s} \rvert \cong \lvert \DGammas \rvert$.
If one inserts Eq.~\eqref{eq:DGamma_sd} into Eq.~\eqref{eq:asl_relation},
one gets the small value of $a^{s}_{sl}$ mentioned before.
These mixing parameters can be measured independently.
In particular, $\Delta m_s$ has already been well-measured.
The current world average is~\cite{HFAG:2010}
%% short-distance \DGamma
\begin{equation}
\Delta m_s   = (17.78\pm 0.12)\ \text{ps}^{-1},
\label{eq:DM_result}
\end{equation}
which is consistent with the SM prediction.
Using the experimental $\Delta m_s$ and $a^s_{sl}$,
Eq.~\eqref{eq:asl_relation} shows that $\Gamma_{12,s}$ has to be enhanced
by at least 3 times of the SM prediction.
In fact,
one of us has already pointed this out~\cite{Hou:2007ps} in 2007,
based on the earlier result of D0,
which has almost the same central value as Ref.~\cite{Abazov:2010hv}
but with larger uncertainty.
Recent studies~\cite{Ligeti:2010ia, Deshpande:2010hy} also indicate this
problem.
On the other hand,
$\DGamma_s$ and $\phi_s$ can also be measured in several ways,
although the precision is not as good as $\Delta m_s$.
One method to extract these values is to study
the $B_s \to J/\psi \phi$ decay.
D0~\cite{D0:2010conf} reported
%% DGamma, D0
\begin{equation}
\begin{split}
\DGamma_s &            = +0.15 \pm 0.06\; \text{(stat)} \pm 0.01\; \text{(syst)}
\,\text{ps}^{-1}, \\
\phi_s &            = -0.76^{+0.38}_{-0.36}\; \text{(stat)}\pm 0.02\; \text{(syst)},
   \label{eq:DGamma_result_D0}
\end{split}
\end{equation}
using $6.1\; \text{fb}^{-1}$ of data.
The consistency of data between mixing parameters
($\Delta m_s$, $\DGammas$, and $\phi_s$) and $a^{s}_{sl}$
has been observed~\cite{Ligeti:2010ia, HFAG:2010}.
Using almost the same amount of data,
CDF~\cite{CDF:2010ICHEP} assumes $\phi_s=0$ and reported
%% DGamma, D0
\begin{equation}
\DGamma_s            = +0.075 \pm 0.035\; \text{(stat)} \pm 0.01\; \text{(syst)}
\,\text{ps}^{-1}
   \label{eq:DGamma_result_CDF}
\end{equation}
This central value drops to half the D0 result,
even below the SM prediction.
But the two results still agree with each other
because the uncertainties so far are still large.
The consistency hints that
new physics may play a role in $B_s-\Bbar_s$ mixing.
New physics can easily enter
the dispersive $M_{12,s}$ and the phase $\phi_s$.
On the other hand,
$\Gamma_{12,s}$ is absorptive and thus hardly affected by new physics at
high energy scale.
As very many properties of $B$ mesons have been studied and found to agree with SM predictions,
new physics has to be rather exotic
to change $\Gamma_{12,s}$ while not affecting other known properties appreciably.

\begin{figure}[t]
\centering
\includegraphics[width=12cm]{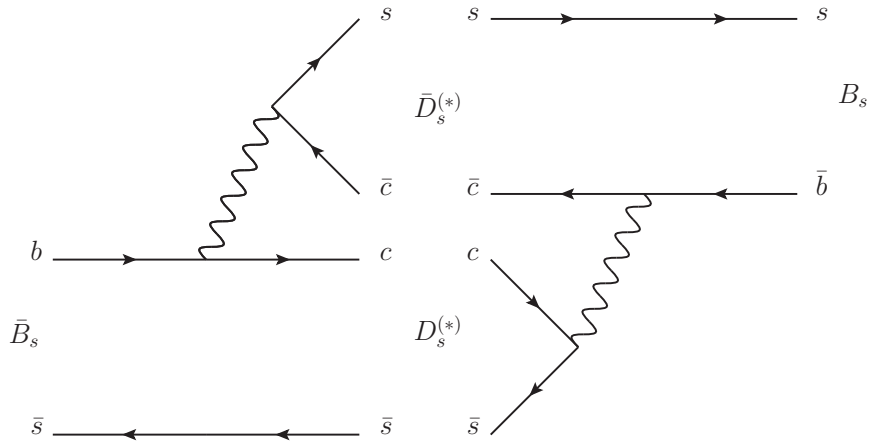}
\caption{The diagrams of $B_s$ and $\Bbar_s$ decay to
$D^{(*)}_s\Dbar^{(*)}_s$ modes.
}
\label{fig:DD_feyndiagram}
\end{figure}

The absorptive nature of $\Gamma_{12,s}$ also makes the
theoretical calculation challenging.
It is helpful to revisit the calculation of $\Gamma_{12,s}$ in SM.
One either approximates $\DGammas$
by operator product expansion (OPE) in short-distance picture, or
estimates $\DGammas$ from several modes which are believed to be important.
The SM prediction~\cite{Lenz:2007JHEP} mentioned previously
adopts the short-distance scheme.
On the other hand, Aleksan \emph{et al.}~\cite{Aleksan:1993qp}
estimated $\DGammas$ from exclusive two-body decays,
mainly $D^{(*)}_s\Dbar^{(*)}_s$ modes
through color-allowed diagrams, as depicted in Fig.~\ref{fig:DD_feyndiagram}.
Their result is close to the current SM prediction.
They further pointed out that $\DGammas$ induced by
$D^{(*)}_s\Dbar^{(*)}_s$ modes
approaches the result of parton model when the limits
$(m_b-2m_c) \to 0$, $m_c \to \infty$ and the large $N_c$ limit
are simultaneously imposed
(for a detail discussion, see Ref.~\cite{Dunietz:2000cr}).
How well does such an approximation hold in Nature remains to be checked.
For example, as Ref.~\cite{Lenz:2007JHEP} and one of us~\cite{Hou:2007ps}
have already pointed out,
a $100\%$ long-distance correction is possible.
The large $a^s_{sl}$ therefore motivates one to investigate the long-distance effect.
In this paper, we perform a detail estimation of
$\DGammas$ from hadronic modes, which includes the two-body modes
$D^{(*,**)}_s \bar D^{(*,**)}_s$, $D^{(*,**)}\bar D_{sJ}(2700)$,
and the three-body $D^{(*)}\Dbar^{(*)}\Kbar^{(*)}$ modes.~\footnote
{Throughout this work, we use $D^{**}_s$ to denote $D_{s0}^*(2317)$, $D_{s1}(2460)$, or $D_{s1}(2536)$.}

%%%%%%%%%%%%%%%%%%%%%%%%%%%%%%%%%%%%
%% DD diagram
%%%%%%%%%%%%%%%%%%%%%%%%%%%%%%%%%%%%%

\begin{figure*}[t]
\centering
\includegraphics[width=12cm]{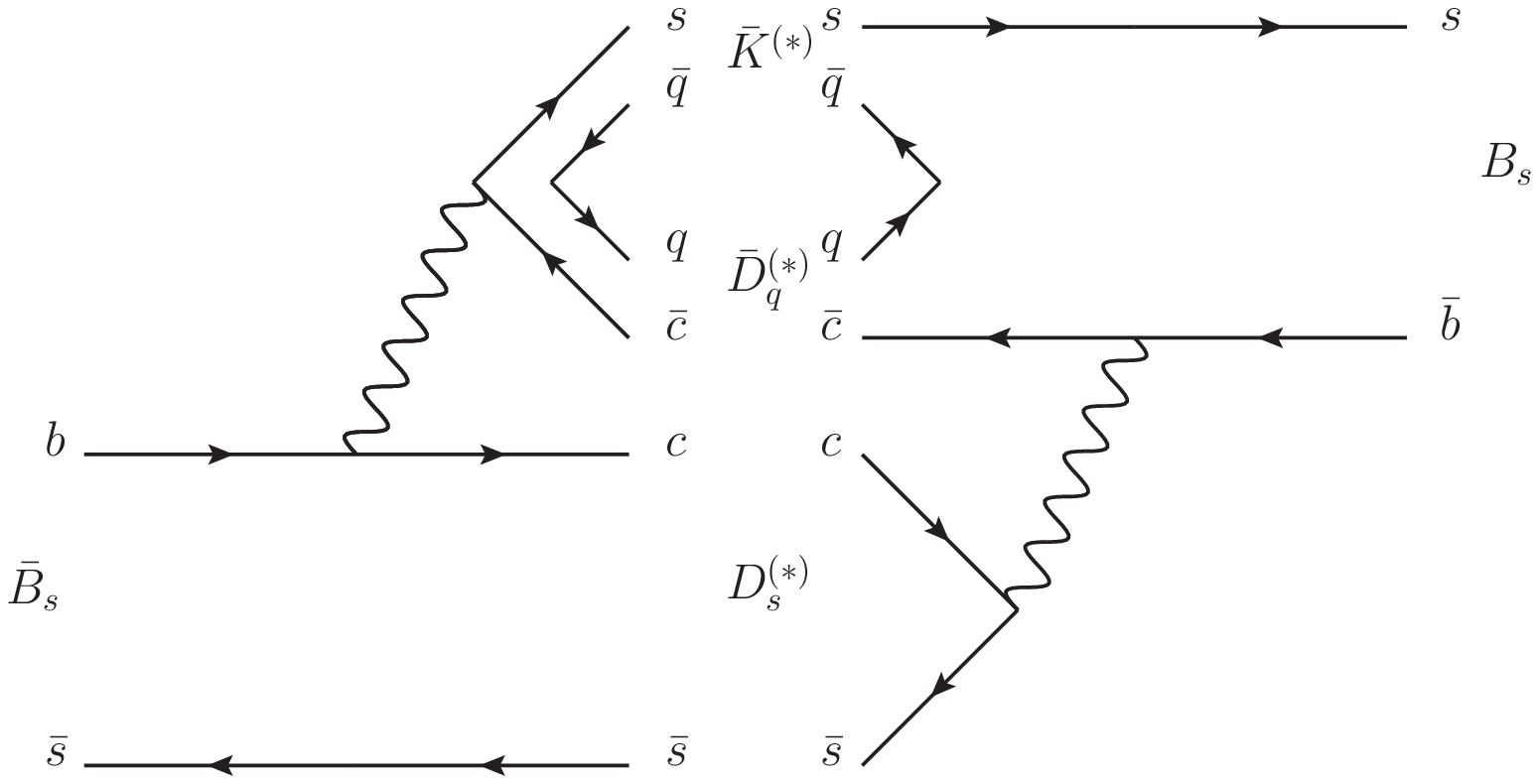}
\caption{The current and transition diagrams.
The left part, $\Bbar_s \to \DDKgeneral$, is the current-produced
diagram, and
the right part, $B_s \to \DDKgeneral$, is transition diagram.}
\label{fig:DDK_feyndiagram}
\end{figure*}

We give the first estimation of the contribution
to $\DGammas$ by three-body $\DDKgeneral$ modes
(and their $\CP$ conjugates).
We use factorization approach, which seems to work well in color-allowed
charmful three-body decays~\cite{Chua:2002pi}, in our calculations.
As shown in Fig.~\ref{fig:DDK_feyndiagram}, these modes can be produced by
the diagram in Fig.~\ref{fig:DD_feyndiagram}, but with an extra $q\bar{q}$
pair produced either in the current or in the spectator part,
which we denote as current-produced ($\mathcal{J}$) or
transition ($\mathcal{T}$) modes.
The number of $D^{(*)}_s\Dbar^{(*)}\Kbar$
channels are four times larger than $D^{(*)}_s\Dbar^{(*)}_s$ modes,
with a factor of two coming from extra $q\bar{q}$,
which can be $u\bar{u}$ or $d\bar{d}$, and another two from
the choice of $q\bar{q}$ in either current or transition processes.
With this enhancement in number of modes,
and if the branching fractions of these modes are not very small compared with
$D^{(*)}_s\Dbar^{(*)}_s$ modes,
it is natural to expect
that $\DGammas$ may receive non-negligible contributions from
three-body $\DDKgeneral$ modes.
So far, the available measurements on these three-body modes
are limited to current-produced modes with $\Kbar$ in $\bar B_{u,d}$
systems only~\cite{Aubert:2003jq, Aubert:2006fh, Brodzicka:2007aa, Aubert:2007rva,
delAmoSanchez:2010pg}.
These modes are related to the corresponding modes in
$\bar B_s$ system under SU(3) symmetry.
We need to reproduce existing three-body data, before we make
predictions for the $\bar B_s$ modes.

Let us briefly survey the experimental situation regarding the
SU(3)-related three-body modes.
There is no measurement of either transition modes or modes with $\Kbar^*$.
Despite a $2.2\sigma$ discrepancy on the branching fraction of
$B^- \to D^{0}\bar D^{0}K^-$ decay between measurements
~\cite{Brodzicka:2007aa, delAmoSanchez:2010pg},
the branching fractions of current-produced
$\Bbar_{u,d} \to D^{(*)}_{u,d}\bar D^{(*)}\bar K$ modes are
around $10^{-2}$ to $10^{-3}$,
one order of magnitude smaller compared to two-body modes.
So far, $c\bar{s}$ resonances $\bar D_{s1}(2536)$ and $\bar D_{sJ}(2700)$
have been observed in the decays $\Bbar_{u,d} \to D^{(*)}_{u,d}\bar D^{(*)}\bar K$
~\cite{Aubert:2006fh, Brodzicka:2007aa, Aubert:2007rva}.
For $D_{s1}(2536)$ resonance,
its contribution to the branching fractions of three-body decays
is in the order of $10^{-4}$, which is small compared with the total branching
fraction.
On the other hand,
Belle observed that $\bar D_{sJ}(2700)$ contributes to about half of the total
branching fraction of $B^- \to D^0 \Dbar^{0}K^-$.
Note that $D_{sJ}(2700)$ has a fairly broad width ($\sim 0.1\,\text{GeV}$).
These measurements suggest that $D_{s1}(2536)$ could be treated in a two-body
picture while it is more appropriate to consider $D_{sJ}(2700)$ in
three-body decays.
Furthermore, the contribution of $\bar D_{sJ}(2700)$ in $B^- \to D^{0}\Dbar^{0}K^-$ decay
is
$\Br(B^- \to D^0 \bar D_{sJ}(2700))\times\Br(\bar D_{sJ}(2700) \to \bar D \bar K)
=(0.113^{+0.026}_{-0.040})\%$,
which is about half the total branching fraction $(0.222\pm0.033)\%$
~\cite{Brodzicka:2007aa}.
Consequently, the contribution of $\bar D_{sJ}(2700)$ in three-body modes
and in $\Delta \Gamma_s$ should be investigated.

This paper is organized as follows.
In Section~\ref{sec:Formalism},
we describe our formalism and briefly review the newly discovered
$D_{sJ}(2700)$ resonance
that has a non-negligible contribution to three-body modes.
The results of two-body modes are in Sec.~\ref{sec:twobody_results}.
For three-body modes,
we examine the factorization formalism and calculate $\DGammas$
in Sec.~\ref{sec:threebody_results}.
Another scenario and the effect of
four-body modes are discussed in Sec.~IV, followed by the concluding section.
Numerical inputs and some calculational details are collected in three Appendices.

%%%%%%%%%%%%%%%%%%%%%%%
\section{Formalism}
\label{sec:Formalism}
%%%%%%%%%%%%%%%%%%%%%%%
\subsection{Formula for $\Delta \Gamma$}
\label{subsec:upperbound}
The time evolution of a $B_s$ meson can be described by the following formula,
%% Time evolution
\begin{equation}
  i \frac{d}{dt}
  \left( {\begin{array}{cc}
      \lvert B \rangle \\
      \lvert \Bbar \rangle
      \end{array} }
  \right) =
  \left(M-i\frac{\Gamma}{2}\right)
  \left( {\begin{array}{cc}
      \lvert B \rangle \\
      \lvert \Bbar \rangle
      \end{array} }
  \right),
\label{eq:eff_h}
\end{equation}
in which we adopt the phase convention of
$\lvert B \rangle$ and $\lvert \Bbar \rangle$ to
be $\CP \lvert B \rangle = - \lvert \Bbar \rangle$.~\footnote
{Our phase convention differs from that in Ref.~\cite{Abazov:2010hv}.}
The $\Gamma$ term in Eq.~(\ref{eq:eff_h}) is the absorptive part,
which can be calculated by summing all on-shell intermediate states,
%% Gamma ab general
\begin{equation}
  \Gamma_{ij} = \frac{1}{2M_B}\sum_{f}\int d\Phi{\mathcal A}^*_{B_i \to f}(\Phi)
                     {\mathcal A}_{B_j \to f}(\Phi),
\end{equation}
where $\Phi$ is over phase space variables.~\footnote{For n-particle mode,
the phase space measure is
$d\Phi=\prod_{j=1}^{n}\frac{dp^3_j}{2E_j}\times
(2\pi)^4\delta^4(\sideset{}{_{j=1}^{n}}\sum p_j-p_B)	
$.}

We define the width difference $\DGammas$ as the difference
between light and heavy eigenstates, $\Gamma_{L}-\Gamma_{H}$.
Assuming $CP$ conservation, which is a good approximation for SM
in $B_s-\Bbar_s$ system,
the eigenstates of $B_s$ meson are $CP$ even and odd states.
From short-distance calculation of SM, the light and heavy eigenstates
correspond to $\CP$ even and odd states respectively.
Thus, the $\DGammas$ can be related to $\Gamma_{ij}$ by
%% \DGammas
\begin{equation}
\begin{split}
  \DGamma \equiv & \ \Gamma_{L}-\Gamma_{H} \\
= & -2 \Gamma_{12} \\
= & -2 \times
    \frac{1}{2M_{B}} \sum_{f} \int d\Phi{\mathcal A}^*_{B \to f}(\Phi)
         {\mathcal A}_{\Bbar \to f}(\Phi),
\end{split}
\label{eq:DGamma_def}
\end{equation}
in which we have used $\Gamma_{21}=\Gamma_{12}^*$ from $CPT$ symmetry, and
$\Gamma_{12}=\Gamma_{12}^*=\text{Re}(\Gamma_{12})$ from $CP$ symmetry.
The fact that $\Gamma_{12}$ is real under $CP$ symmetry can be seen from
%% real Gamma12
\begin{equation}
\begin{split}
    \Gamma_{12} & =
          \frac{1}{2M_{B}} \sum_{f} \int d\Phi{\mathcal A}^*_{B \to f}(\Phi)
          {\mathcal A}_{\Bbar \to f}(\Phi) \\
     & =  \frac{1}{2M_{B}} \sum_{f} \frac{1}{2} \int d\Phi
                ({\mathcal A}^*_{B \to f}(\Phi)
                 {\mathcal A}_{\Bbar \to f}(\Phi) +
                 {\mathcal A}^*_{B \to \bar{f}}(\Phi)
                 {\mathcal A}_{\Bbar \to \bar{f}}(\Phi)) \\
     & =  \frac{1}{2M_{B}} \sum_{f} \text{Re}\left[
                \int d\Phi{\mathcal A}^*_{B \to f}(\Phi)
                     {\mathcal A}_{\Bbar \to f}(\Phi)\right].
\end{split}
\label{eq:real_Gamma12}
\end{equation}
The amplitude product
${\mathcal A}^*_{B \to f}(\Phi){\mathcal A}_{\Bbar \to f}(\Phi)$ is complex
conjugate to the amplitude product of conjugate intermediate state
${\mathcal A}^*_{B \to \bar{f}}(\Phi){\mathcal A}_{\Bbar \to \bar{f}}(\Phi)$ by
$CP$ symmetry. $\Gamma_{12}$ sums up all the intermediate states and turns out
to be real. For convenience, we define the width difference
of each exclusive decay as $\DGamma_f$, and its corresponding
complex term in $\Gamma_{12}$ to be
%% DGammaf and Gamma12f
\begin{equation}
\begin{split}
    \DGamma_{f} & \equiv
-2 \times \text{Re}[\Gamma_{12,f}],
\end{split}
\label{eq:DGammaf}
\end{equation}
where $\Gamma_{12,f}$ is defined as
%% Gamma12f
\begin{equation}
\begin{split}
    \Gamma_{12,f} \equiv
          \frac{1}{2M_{B}} \int d\Phi{\mathcal A}^*_{B \to f}(\Phi)
          {\mathcal A}_{\Bbar \to f}(\Phi).
\end{split}
\label{eq:Gamma12f}
\end{equation}
Although $\Gamma_{12,f}$ is complex by looking at one mode,
the imaginary part is cancelled by its $\CP$ conjugate mode, and thus
the total $\Gamma_{12}$ turns out to be real.
Once ${\mathcal A}_{B \to f}(\Phi)$ and ${\mathcal A}_{\bar{B} \to f}(\Phi)$
are known, one can readily calculate the corresponding $\DGamma_f$ and
branching fractions. In the next section, we will apply the
factorization formalism to obtain these amplitudes.

Before we move to  model-dependent calculation, it is useful to extract some
general limits of the magnitude of $\DGamma_f$
from Eq.~(\ref{eq:DGammaf}).
For an intermediate state $\lvert f \rangle$, the magnitude of
$\DGamma_f$ induced by this state is bounded by
%% DGamma limit
\begin{subequations} \label{eq:DGamma_limit}
 \begin{align}
   \left|\frac{\DGamma_f}{\Gamma} \right| = & \
     \frac{1}{\Gamma} \cdot  \left| \text{Re}[2 \Gamma_{12,f}] \right| \\
     \leq & \ \frac{1}{\Gamma} \cdot  \left| 2 \Gamma_{12,f}\right|
\label{eq:DGamma_limit_1}\\
     \leq & \ \frac{2}{\Gamma} \cdot \frac{1}{2M_{B}}
            \int d\Phi \sqrt{|{\mathcal A}_{\bar{B} \to f}(\Phi) |^2}
            \sqrt{| {\mathcal A}_{B \to f}(\Phi) | ^2}
\label{eq:DGamma_limit_2}\\
     \leq & \ 2 \times \sqrt{{\mathcal B}_{\bar{B} \to f}} \times
            \sqrt{{\mathcal B}_{B \to f}}.
\label{eq:DGamma_limit_3}
% \end{subequations}
\end{align}
\end{subequations}
There are three inequalities in this formula.
The first inequality reflects that $\DGamma_f$ is only proportional
to the real part of $\Gamma_{12,f}$.
The second inequality is obtained by the fact that
the phase of the amplitude product
${\mathcal A}^*_{B \to f}(\Phi){\mathcal A}_{\Bbar \to f}(\Phi)$
may be different over the phase space,
which would reduce the overall $\lvert \Gamma_{12,f} \rvert$.
For the last inequality, it accounts for the ``mismatch'' effect between
$|{\mathcal A}_{B \to f}(\Phi)|$ and $|{\mathcal A}_{\bar{B} \to f}(\Phi)|$.
Even when the branching fractions of $\Bbar \to f$ and $B \to f$ are the same,
the induced $\Delta \Gamma$ could be quite small if the decay probabilities of
the two modes are highly mismatched in phase space.
Note that the latter two limits are experimental observables.
If the branching fractions of $\Bbar \to f$ and $B \to f$ are measured,
one could find the maximal magnitude of the corresponding $\Gamma_{12,f}$.
The bound can be refined by the second inequality
if the Dalitz plots of the two modes are available.
But the $\DGamma_f$, which is proportional to the real part $\Gamma_{12,f}$,
could be any value in the range of $-2\lvert \Gamma_{12,f} \rvert$
to $+ \lvert 2\Gamma_{12,f} \rvert$.

\subsection{Factorization Formalism}

The relevant effective Hamiltonian for the
$b\to c$ transition is
\begin{equation}
{\mathcal H_{\rm eff}}=
\frac{G_F}{\sqrt2} V_{cb} V_{cs}^*
\big[
   c_1(\mu)\mathcal O^c_1(\mu)+c_2(\mu)\mathcal O^c_2(\mu)\big],
 \label{eq:Heff}
\end{equation}
where $c_i(\mu)$ are the Wilson coefficients, and $V_{cb}$
and $V_{cs}$ are the Cabibbo-Kobayashi-Maskawa~(CKM) matrix elements.
The four-quark operators ${\mathcal O}_i$ are products of two
$V-A$ currents, i.e.
${\mathcal O^c_1}=(\bar c b)_{V-A}\,(\bar s c)_{V-A}$ and
${\mathcal O^c_2}=(\bar s b)_{V-A}\,(\bar c c)_{V-A}$.

With the factorization ansatz, the amplitudes for
two-body $\Bbar_{s} \to \mathcal{D}^{(*)}_s \mathcal{\bar{D}}^{(*)}_s$
decays are given by
%% DD amp
\begin{equation}
{\mathcal A}(\Bbar_{s} \to \mathcal{D}^{(*)}_s \mathcal{\bar{D}}^{(*)}_s)=
\frac{G_F}{\sqrt2} V_{cb} V_{cs}^*
a_1 \langle \mathcal{D}^{(*)}_s |(\bar c b)_{V-A}|\Bbar_{s}\rangle
    \langle \mathcal{\bar{D}}^{(*)}_s|(\bar s c)_{V-A}|0\rangle,
 \label{eq:DDAmp}
\end{equation}
where the effective coefficients are expressed as
$a_1=c_1+c_2/3$ if naive factorization is used.
Note that $\mathcal {D}^{(*)}_s$ could be the usual $D^{(*)}_s$ and
or higher $D_s$ resonance such as $D_{s0}^*(2317)$, $D_{s1}(2460)$,
$D_{s1}(2536)$, and $D_{sJ}(2700)$.
The factorized amplitudes consist of the products of two common
matrix elements: the current-produced $\mathcal{D}^{(*)}_s$ and
the $\Bbar_s$ to $\mathcal{D}^{(*)}_s$ transition.
They are parametrized by the standard way~\cite{Cheng:2003sm}.
The matrix elements of current-produced $\mathcal{D}_s^{(*)}$ are
%% eq: D, Dst decay constants
\begin{equation}
\begin{split}
\langle \mathcal{D}_s(p)|(V-A)_{\mu}|0\rangle = &
   \quad if_{\mathcal{D}_s}p_{\mu}, \\
\langle \mathcal{D}^*_s(p, \lambda)|(V-A)_{\mu}|0\rangle = &
   \quad m_{\mathcal{D}^*_s}f_{\mathcal{D}^*_s}\eps^{*}_{\mu}(p, \lambda).
\end{split}
 \label{eq:D_decayconst}
\end{equation}
The transition matrix elements for $\mathcal{D}_s^{(*)}$ are
%% eq: B to D(*) Transition Form Factors
\begin{equation}
\begin{split}
  \langle \mathcal{D}_s(p_D)|(V-A)_{\mu}|\Bbar_s(p_B)\rangle =
&
  \quad \left((p_B+p_D)_{\mu}-\frac{m_{B}^2-m_{D}^2}{q^2}q_{\mu}\right)
    F_1^{\Bbar_s \mathcal{D}_s}(q^2)
  +\frac{m_{B}^2-m_{D}^2}{q^2}q_{\mu} F_0^{\Bbar_s \mathcal{D}_s}(q^2),
  \\
\langle \mathcal{D}^*_s(p_{\Dst}, \lambda)|(V-A)_{\mu}|\Bbar_s(p_B)\rangle =
&
\quad \epsilon_{\mu \nu \rho \sigma} \eps^{*\nu} p_B^{\rho} p_{\Dst}^{\sigma}
    \cdot \frac{2F_3^{\Bbar_s \mathcal{D}^*_s}(q^2)}{m_B+m_{\Dst}}
 \\
& -i\left(\eps^{*}_{\mu}-\frac{\eps^{*}\cdot q}{q^2}q_{\mu}\right)
    \,(m_B+m_{\Dst})F_1^{\Bbar_s \mathcal{D}^*_s}(q^2)
 \\
& +i\left((p_B+p_{\Dst})_{\mu}-\frac{m_{B}^2-m_{\Dst}^2}{q^2}q_{\mu}\right)(\eps^{*} \cdot q)
    \, \frac{F_2^{\Bbar_s \mathcal{D}^*_s}(q^2)}{m_B+m_{\Dst}}
 \\
& -i\frac{\eps^{*}\cdot q}{q^2}q_{\mu}
    \, 2 m_{\Dst} F_0^{\Bbar_s \mathcal{D}^*_s}(q^2),
\end{split}
 \label{eq:BD_transform}
\end{equation}
where $\epsilon_{\mu \nu \rho \sigma}$ is the totally anti-symmetric symbol with
$\epsilon_{0123}=1$.
For convenience, our notations of decay constants and
form factors of $D^{**}_s$ are different from the usual notations.
The conversion can be found in Appendix~\ref{FF_transform}.

The amplitudes of three-body modes
$D^{(*)} \Dbar^{(*)} \Kbar^{(*)}$ decayed from $\Bbar$ and $B$
are given by
\begin{equation}
\begin{split}
{\mathcal A_{\Cur}}(\Bbar_s \to D_s^{(*)}\Dbar^{(*)} \Kbar^{(*)}) & =
\frac{G_F}{\sqrt2} V_{cb} V_{cs}^*
a_1 \langle D_s^{(*)}|(\bar c b)_{V-A}|\Bbar_s \rangle \cdot
    \langle \Dbar^{(*)}\Kbar^{(*)}|(\bar s c)_{V-A}|0\rangle, \\
{\mathcal A_{\Tr}}(B_s \to D_s^{(*)}\Dbar^{(*)} \Kbar^{(*)}) & =
\frac{G_F}{\sqrt2} V_{cb} V_{cs}^*
a_1 \langle \Dbar^{(*)}\Kbar^{(*)}|(\bar c b)_{V-A}|B_s \rangle \cdot
    \langle D_s^{(*)} |(\bar s c)_{V-A}|0\rangle,
\end{split}
 \label{eq:DDKAmp}
\end{equation}
where $\mathcal A_{\Cur}$ and $\mathcal A_{\Tr}$ denote the amplitudes of
current and transition diagrams, respectively.
Unlike the $\mathcal{D}^{(*)}_s \mathcal{\bar{D}}^{(*)}_s$ modes in which
only standard form factors appear,
these amplitudes involve the time-like form factors and transition form factors
of two pseudoscalars ($\bar D\bar K$) or vectors ($\bar D^*\bar K^*$),
or a pseudoscalar with a vector ($\bar D^*\bar K$ or $\bar D\bar K^*$).

The parametrization of time-like form factors are similar to the space-like
counterparts, such as $\langle D_s^{(*)}|V-A|\Bbar_s \rangle$.
The time-like form factors of two pseudoscalars ($PP$) states are given by
%% eq: PP Time-like Form Factors
\begin{equation}
\begin{split}
\langle P_a(p_a)P_b(p_b)|(V-A)_{\mu}|0\rangle = &
 \left((p_a-p_b)_{\mu}-\frac{m_{a}^2-m_{b}^2}{q^2}q_{\mu}\right) F_1^{PP}(q^2)
 +\frac{m_{a}^2-m_{b}^2}{q^2}q_{\mu} F_0^{PP}(q^2),
\end{split}
 \label{eq:PP_Cur_Form}
\end{equation}
where $q^{\mu}=p_a^{\mu}+p_b^{\mu}$ is the momentum of the current.
For the states with one vector and pseudoscalar ($VP$), the parametrization of
time-like form factors are
%% eq: VP Time-like Form Factors
\begin{equation}
\begin{split}
\langle V(p_V, \eps_V)P(p_P)|(V-A)_{\mu}|0\rangle =  &
 -\epsilon_{\mu \nu \rho \sigma} \eps_V^{*\nu} p_P^{\rho} p_V^{\sigma}
    \cdot \frac{2V^{VP}(q^2)}{m_V+m_P}
 -i\left(\eps^{*}_{V\mu}-\frac{\eps_V^{*}\cdot q}{q^2}q_{\mu}\right)
    (m_V+m_P)A_1^{VP}(q^2)\\
& -i\left((p_V-p_P)_{\mu}-\frac{m_{V}^2-m_{P}^2}{q^2}q_{\mu}\right)(\eps_V^{*} \cdot q)
    \, \frac{A_2^{VP}(q^2)}{m_V+m_P}
 -i\frac{\eps_V^{*}\cdot q}{q^2}q_{\mu}
    \, 2 m_V A_0^{VP}(q^2).
\end{split}
 \label{eq:VP_Cur_Form}
\end{equation}
The time-like form factors of two vectors ($VV$) states can be
parameterized analogously,
%% eq: VV Time-like Form Factors
\begin{equation}
\begin{split}
\langle V_a(p_a, \eps_a)V_b(p_b, \eps_b)|(V-A)_{\mu}|0\rangle =  &
 \quad i\epsilon_{\alpha \nu \rho \sigma} \eps_b^{*\alpha}\eps_a^{*\nu} p_a^{\rho} p_b^{\sigma}q_{\mu}
    \frac{V^{VV}_0(q^2)}{(m_a+m_b)^2}
 +i\epsilon_{\mu \nu \rho \sigma} \eps_a^{*\nu} p_a^{\rho} p_b^{\sigma}
    (\eps_b^* \cdot q)
    \frac{V^{VV}_1(q^2)}{(m_a+m_b)^2} \\
& +i\epsilon_{\mu \nu \rho \sigma} \eps_b^{*\nu} p_a^{\rho} p_b^{\sigma}
    (\eps_a^* \cdot q)
     \frac{V^{VV}_2(q^2)}{(m_a+m_b)^2}
 +\left(\eps^{*}_{a\mu}-\frac{\eps_a^{*}q}{q^2}q_{\mu}\right)(\eps_b^{*} \cdot q)
     A_{11}^{VV}(q^2)\\
& +\left(\eps^{*}_{b\mu}-\frac{\eps_b^{*} \cdot q}{q^2}q_{\mu}\right)(\eps_a^{*} \cdot q)
     A_{12}^{VV}(q^2)
 +\left((p_a-p_b)_{\mu}-\frac{m_{a}^2-m_{b}^2}{q^2}q_{\mu}\right)(\eps_a^{*} \cdot \eps_b^{*})
     A_2^{VV}(q^2)\\
& +(\eps_a^{*} \cdot q)(\eps_b^{*} \cdot q)\frac{q_{\mu}}{q^2}
     A_{01}^{VV}(q^2),
 +(\eps_a^{*} \cdot \eps_b^{*})\frac{q_{\mu}}{q^2}
     (m_a+m_b)^2 A_{02}^{VV}(q^2).
\end{split}
 \label{eq:VV_Cur_Form}
\end{equation}

The transition form factors are more complicated.
The case of $B_s$ to $PP$ transition form factors were formulated
in a general way in Ref.~\cite{Lee:ih}, which can be rewritten as
%% eq: PP Transition Form Factors
\begin{equation}
\begin{split}
\langle P_a(p_a)P_b(p_b)|(V-A)_{\mu}|\Bbar_s(p_B) \rangle = &
\quad \epsilon_{\mu \nu \rho \sigma} p_B^{\nu} q^{\rho} (p_a-p_b)^{\sigma}
    \, \frac{V^{\Bbar_s PP} }{m_{B_s}^3}
  +i\left((p_B+q)_{\mu}-\frac{m_{B_s}^2-q^2}{q'^2}q'_{\mu}\right)
    \, \frac{A_1^{\Bbar_s PP} }{m_{B_s}}\\
& +i\left((p_a-p_b)_{\mu}-\frac{m_{a}^2-m_{b}^2}{q^2}q_{\mu}\right)
    \, \frac{A_2^{\Bbar_s PP} }{m_{B_s}}
  +i\frac{m_a^2-m_b^2}{q^2}q_{\mu}
    \, \frac{A_0^{\Bbar_s PP} }{m_{B_s}},
\end{split}
 \label{eq:PP_Tr_Form}
\end{equation}
where $q^{\mu}=p_a^{\mu}+p_b^{\mu}$ is the total momentum of $PP$, and
$q'^{\mu}=p_B^{\mu}-q^{\mu}$ is the momentum of the external current.
In this form, the terms with $A_1$ and $A_2$
are zeros when contracted with $q'$ and $q$.
For the transition form factors of $\Bbar_s$ to $VP$ and $VV$,
since they are more complicated and there is so far no data, we only write down
the form factors obtained from pole model rather than the general forms.
For $VP$, we have
%% eq: VP Transition Form Factors
\begin{equation}
\begin{split}
\langle V(p_V, \eps_V)P(p_P)|(V-A)^{\mu}|\Bbar_s(p_B) \rangle = &
   \quad i\epsilon_{\alpha \nu \rho \sigma}
    \left(-g^{\mu \alpha}+\frac{q'^{\mu}q'^{\alpha}}{q'^2}\right)\eps_V^{*\nu} p_P^{\rho} p_V^{\sigma}
    \, \frac{V^{\Bbar_s VP}_{2} }{m_{B_s}^2}\\
&  +i\epsilon_{\alpha \nu \rho \sigma} q'^{\alpha}\eps_V^{*\nu} p_P^{\rho} p_V^{\sigma}
    \left((p_B+q)^{\mu}-\frac{m_{B_s}^2-q^2}{q'^2}q'^{\mu}\right)
    \, \frac{V^{\Bbar_s VP}_{1} }{m_{B_s}^4}\\
&  +i\epsilon_{\alpha \nu \rho \sigma} q'^{\alpha}\eps_V^{*\nu} p_P^{\rho} p_V^{\sigma}
    \frac{q'^{\mu}}{q'^2}
    \, \frac{V^{\Bbar_s VP}_{0} }{m_{B_s}^2}\\
&  +\epsilon_{\alpha \beta \gamma \delta} \epsilon_{abcd}
    (g^{\mu \alpha}g^{\beta a})q'^{\gamma}q^{\delta}
    \eps_V^{*b} p_P^{c} p_V^{d}
    \, \frac{A^{\Bbar_s VP}_{3} }{m_{B_s}^4}\\
&  +\left((p_B+q)^{\mu}-\frac{m_{B_s}^2-q^2}{q'^2}q'^{\mu}\right)
    (\eps_V^{*} \cdot q)
    \, \frac{A_1^{\Bbar_s VP} }{m_{B_s}^2}
   +\frac{m_{B_s}^2-q^2}{q'^2} q'^{\mu}
    (\eps_V^{*} \cdot q)
    \, \frac{A_0^{\Bbar_s VP} }{m_{B_s}^2},
\end{split}
 \label{eq:VP_Tr_Form}
\end{equation}
and for $\Bbar_s$ to $VV$, we parameterize as
%% eq: VV Transition Form Factors
\begin{equation}
\begin{split}
 & \langle V_a(p_a, \eps_a)V_b(p_b)|(V-A)_{\mu}|\Bbar_s(p_B) \rangle  \\
 & \ \ \ \ \ \ \ \ \ \ \ \ \ = \epsilon_{\mu \nu \rho \sigma}
    p_a^{\nu}p_b^{\rho} q'^{\sigma} (\eps_a^{*} \cdot \eps_b^{*})
    \, \frac{V^{\Bbar_s VV}_{3} }{m_{B_s}^3}
  +\epsilon_{\mu \nu \rho \sigma} \eps_a^{*\nu} q'^{\rho} q^{\sigma}
    (\eps_b^{*} \cdot q)
    \, \frac{V^{\Bbar_s VV}_{2} }{m_{B_s}^3}\\
 & \ \ \ \ \ \ \ \ \ \ \ \ \ \ \  +\epsilon_{\mu \nu \rho \sigma} \eps_b^{*\nu} q'^{\rho} q^{\sigma}
    (\eps_a^{*} \cdot q)
    \, \frac{V^{\Bbar_s VV}_{1} }{m_{B_s}^3}
   +\epsilon_{\alpha \nu \rho \sigma} \eps_b^{*\alpha} \eps_a^{*\nu} p_a^{\rho} p_b^{\sigma}
    \left((p_B+q)_{\mu}-\frac{m_{B_s}^2-q^2}{q'^2}q'_{\mu}\right)
    \, \frac{V^{\Bbar_s VV}_{01} }{m_{B_s}^3}\\
 & \ \ \ \ \ \ \ \ \ \ \ \ \ \ \  +\epsilon_{\alpha \nu \rho \sigma} \eps_b^{*\alpha} \eps_a^{*\nu} p_a^{\rho} p_b^{\sigma}
    \frac{m_{B_s}^2-q^2}{q'^2} q'_{\mu}
    \, \frac{V^{\Bbar_s VV}_{00} }{m_{B_s}^3}\\
 & \ \ \ \ \ \ \ \ \ \ \ \ \ \ \  +i(\eps_a^{*} \cdot \eps_b^{*})
    \left((p_a-p_b)_{\mu}-\frac{m_{a}^2-m_{b}^2}{q'^2}q'_{\mu}\right)
    \, \frac{A_{62}^{\Bbar_s VV} }{m_{B_s}}\\
 & \ \ \ \ \ \ \ \ \ \ \ \ \ \ \  +i(\eps_a^{*} \cdot \eps_b^{*})
    \left((p_B+q)_{\mu}-\frac{m_{B_s}^2-q^2}{q'^2}q'_{\mu}\right)
    \, \frac{A_{61}^{\Bbar_s VV} }{m_{B_s}}
   +i(\eps_a^{*} \cdot \eps_b^{*})
    \frac{q'_{\mu}}{q'^2}
    \, m_{B_s} A_{60}^{\Bbar_s VV} \\
 & \ \ \ \ \ \ \ \ \ \ \ \ \ \ \  +i(\eps_b^{*} \cdot q)
    \left(\eps^{*}_{a\mu}-\frac{\eps_a^{*}q}{q^2}q_{\mu}\right)
    \, \frac{A_{3}^{\Bbar_s VV} }{m_{B_s}}
   +i(\eps_a^{*} \cdot q)
    \left(\eps^{*}_{b\mu}-\frac{\eps_b^{*}q}{q^2}q_{\mu}\right)
    \, \frac{A_{4}^{\Bbar_s VV} }{m_{B_s}}\\
 & \ \ \ \ \ \ \ \ \ \ \ \ \ \ \  +i(\eps_a^{*} \cdot q')(\eps_b^{*} \cdot q)
    \left((p_B+q)_{\mu}-\frac{m_{B_s}^2-q^2}{q'^2}q'_{\mu}\right)
    \, \frac{A_{21}^{\Bbar_s VV} }{m_{B_s}^3}
   +i(\eps_a^{*} \cdot q')(\eps_b^{*} \cdot q)
    \frac{q'_{\mu}}{q'^2}
    \, \frac{A_{20}^{\Bbar_s VV} }{m_{B_s}}\\
 & \ \ \ \ \ \ \ \ \ \ \ \ \ \ \  +i(\eps_a^{*} \cdot q)(\eps_b^{*} \cdot q')
    \left((p_B+q)_{\mu}-\frac{m_{B_s}^2-q^2}{q'^2}q'_{\mu}\right)
    \, \frac{A_{11}^{\Bbar_s VV} }{m_{B_s}^3}
   +i(\eps_a^{*} \cdot q)(\eps_b^{*} \cdot q')
    (\frac{q'_{\mu}}{q'^2})
    \, \frac{A_{10}^{\Bbar_s VV} }{m_{B_s}}\\
 & \ \ \ \ \ \ \ \ \ \ \ \ \ \ \  +i(\eps_a^{*} \cdot q)(\eps_b^{*} \cdot q)
    \left((p_B+q)_{\mu}-\frac{m_{B_s}^2-q^2}{q'^2}q'_{\mu}\right)
    \, \frac{A_{01}^{\Bbar_s VV} }{m_{B_s}^3}
   +i(\eps_a^{*} \cdot q)(\eps_b^{*} \cdot q)
    \frac{q'_{\mu}}{q'^2}
    \, \frac{A_{00}^{\Bbar_s VV} }{m_{B_s}}.\\
\end{split}
 \label{eq:VV_Tr_Form}
\end{equation}

Under $\CP$ conservation, all these form factors can be related to the
form factors of their $\CP$ conjugates. These transformations are provided
in Appendix~\ref{FF_transform}.

\subsection{Pole Model}

%%%%%%%%%%%%%%%%%%%%%%%%%%%%%%%%%%%%%%%%%%%%%%%%%%%%%%%%
\begin{figure}[t]
\centering
\includegraphics[width=14cm]{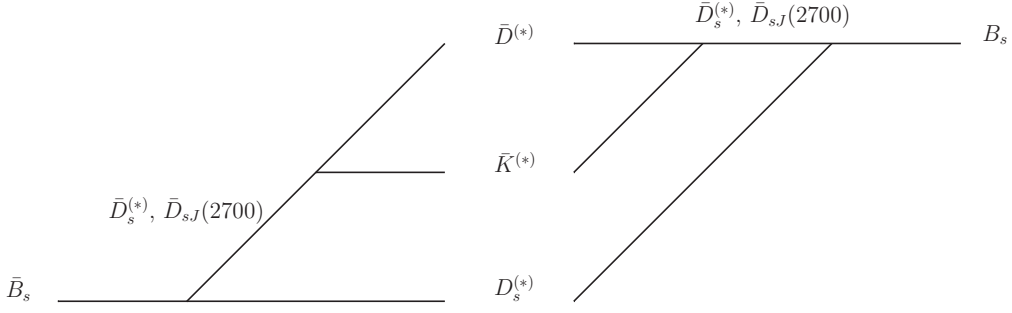}
\caption{Pole diagram of $\Bbar_s-B_s$ mixing.
Left: current-produced diagram.
Right: transition diagram.}
\label{fig:pole_diagram}
\end{figure}
%%%%%%%%%%%%%%%%%%%%%%%%%%%%%%%%%%%%%%%%%%%%%%%%%%%%%%%%

Since the branching fractions of $D^{(*)}_s\Dbar^{(*)}_s$ are large,
it is natural to expect a sizable contribution from off-shell $D^{(*)}_s$ poles.
In addition, experiments have observed
$D_{sJ}(2700)$
in the three-body decays
as we have described in the introduction
~\cite{Aubert:2006fh, Brodzicka:2007aa}.
$D_{sJ}(2700)$ can decay to on-shell $D^{(*)}K$,
but only goes off-shell to $D^{(*)} \Kst$ because of kinematics.
As shown in Fig.~\ref{fig:pole_diagram}, we consider pole exchanges, including $D_s$, $D^*_s$ and $D_{sJ}(2700)$, in three-body decays.
%The pole diagrams of $\Bbar_s-B_s$ mixing,
%which is mediated by current-produced and transition diagrams,
%are shown in Fig.~\ref{fig:pole_diagram}.
Note that the $D_s$ pole only goes to $D^*K$ rather than $DK$.

In the following calculation,
we use off-shell $D^{(*)}_s$ poles and $D_{sJ}(2700)$
to model the $D^{(*)}K^{(*)}$ form factors.
The effective Lagrangian
taken from Ref.~\cite{Casalbuoni:1996pg, Yan:1992gz, Cheng:2004ru}
is applied to describe the interaction between
$D^{(*)}_q$ mesons and light psedudoscalar or vector mesons.
The pole contribution to $D^{(*)}K^{(*)}$ form factors can be calculated by
%% General Form of Pole contribution
\begin{equation}
\begin{split}
\langle D^{(*)}K^{(*)}|(V-A)_{\mu}|0\rangle_{\rm pole} =
&
  \frac{i}{q^2-m_{\rm int}^2 + im_{\rm int}\Gamma_{\rm int}} \times
\langle D^{(*)}K^{(*)}|i \mathcal{L}_{\rm eff} |D_{\rm int}\rangle
   \langle D_{\rm int}|(V-A)_{\mu}|0 \rangle \\
&
 + \frac{i}{q^2-m_{\rm int*}^2 + im_{\rm int*}\Gamma_{\rm int*}} \times
  \left(-g^{\alpha \beta}+\frac{q^{\alpha}q^{\beta}}{m^2_{\rm int*}}\right) \\
& \times\frac{\partial^2}{\partial \eps_{\rm int}^{*\alpha} \partial \eps_{\rm int}^{\beta}}
  \left(\langle D^{(*)}K^{(*)}|i \mathcal{L}_{\rm eff} |D^*_{\rm int}\rangle
   \langle D^*_{\rm int}|(V-A)_{\mu}|0 \rangle\right ),
\\
\langle D^{(*)}K^{(*)}|(V-A)_{\mu}|\Bbar \rangle_{\rm pole} = &
  \frac{i}{q^2-m_{\rm int}^2 + im_{\rm int}\Gamma_{\rm \rm int}} \times
    \langle D^{(*)}K^{(*)}|i \mathcal{L}_{\rm eff} |D_{\rm int}\rangle
   \langle D_{int}|(V-A)_{\mu}|\Bbar \rangle\\
&+
  \frac{i}{q^2-m_{\rm int*}^2 + im_{\rm int*}\Gamma_{\rm \rm int*}} \times
  \left(-g^{\alpha \beta}+\frac{q^{\alpha}q^{\beta}}{m^2_{\rm int*}}\right)
\\
& \times\frac{\partial^2}{\partial \eps_{\rm int}^{*\alpha} \partial \eps_{\rm int}^{\beta}}
  \left(\langle D^{(*)}K^{(*)}|i \mathcal{L}_{\rm eff} |D^*_{\rm int}\rangle
   \langle D^*_{int}|(V-A)_{\mu}|\Bbar \rangle\right),
\end{split}
 \label{eq:pole_derivation}
\end{equation}
where the $D^{(*)}_{\rm int}$ is the intermediate particle with mass $m_{\rm int(*)}$
and width $\Gamma_{\rm int(*)}$. We adopt the Breit-Wigner form of the propagator and
replace $\eps_{\rm int}^{*\alpha} \eps_{\rm int}^{\beta}$ as
$(-g^{\alpha \beta}+{q^{\alpha}q^{\beta}}/{m^2_{\rm int*}})$ to account for the
off-shell effect.
The explicit forms of the matrix elements
$\langle D^{(*)}K^{(*)}|i \mathcal{L}_{\rm eff} |D^{(*)}_{\rm int}\rangle$
in the above equations can be found
in Ref.~\cite{Cheng:2004ru}.
A full list of pole contribution to form factors are listed
in Appendix~\ref{full_pole}.

\subsection{$D_{sJ}(2700)$ Resonance}
\label{sec:DsJ_properties}
The relevant properties and parameters of $D_{sJ}(2700)$
are summarized in this section.
The mass and width of this resonance are~\cite{PDG}
\begin{equation}
\begin{split}
 m_{D_{sJ}(2700)}      & =2709^{+9}_{-6} \text{ MeV},\\
 \Gamma_{D_{sJ}(2700)} & =125\pm30 \text{ MeV}.
\end{split}
 \label{eq:DsJ_masswidth}
\end{equation}
Note that the width has a large uncertainty ($\sim 25\%$).
The ratio of branching fractions of this resonance to $DK$ and $\Dst K$ is
also measured~\cite{Aubert:2009di}
%% DsJ exp. branching fraction fraction
\begin{equation}
 \begin{split}
 r(\Dst K) \equiv
 \frac{\Br(D_{sJ}(2700)^+ \to \Dst K)}{\Br(D_{sJ}(2700)^+ \to DK)} &
    = 0.91\pm0.13_{\rm stat}\pm0.12_{\rm syst},
 \end{split}
 \label{eq:DsJ_br_ratio_exp}
\end{equation}
where $D^{(*)}K$ is the average of $D^{(*)}K_S$ and $D^{(*)}K^+$ modes.
On the other hand,
the contribution of $D_{sJ}(2700)$ in the decay
$B^+ \to \Dbar^{0}D^{0}K^+$, denoted as
$\Br(B^+ \to \Dbar^{0}D_{sJ}(2700))\times\Br(D_{sJ}(2700) \to D^{0}K^+)$,
is extracted~\cite{Brodzicka:2007aa}
%% DsJ exp. branching fraction fraction
\begin{equation}
 \begin{split}
   \Br(B^+ \to \Dbar^0 D_{sJ}(2700))\times\Br(D_{sJ}(2700) \to DK)
   =(11.3^{+2.6}_{-4.0})\times 10^{-4},
 \end{split}
 \label{eq:DsJ_partialbr_exp}
\end{equation}
which constitutes about half the total branching fraction
of this measurement.
Note that this quantity has a large uncertainty,
similar to the measurement of width.
The quantum number of $D_{sJ}(2700)$ is determined to be
$J^P=1^{-}$ from helicity angle distribution, which
limits this resonance to be either an $s$-wave or $d$-wave meson
(or a mixed state between them).
The interpretation of $D_{sJ}(2700)$ as a radial excitation of $\Dst_s$
($n^{2S+1}L_{J}=2^3S_{1}$) is proposed, which can explain
its mass~\cite{Godfrey:1985xj},
partial width~\cite{Colangelo:2007ds},
and contribution in $B^+ \to \Dbar^{0}D^{0}K^+$ decay~\cite{Wang:2009zz}.
In some strong decay models, a mixed state $2^3S_{1}-1^3D_{1}$
describes the partial width better~\cite{Close:2006gr}.
As the theoretical predictions of mass and partial width
are highly model-dependent, the identification is still not clear yet.
We assume $D_{sJ}(2700)$ as a $2^3S_1$ state in this study.

The effective Lagrangian in
Ref.~\cite{Casalbuoni:1996pg, Yan:1992gz, Cheng:2004ru} can still be applied
to describe the interaction between $D_{sJ}(2700)$
and light mesons~\cite{Colangelo:2007ds}.
We work out the relevant matrix elements,
%% DsJ form factors
\begin{equation}
\begin{split}
\langle D(p_2)K(p_3)|i\mathcal{L}_{\rm eff}|D_{sJ}(2700)(p_1,\eps_1)\rangle = &
    -i\tilde{g}_{D_{sJ} DK}\, \eps_1 \cdot p_3,\\
\langle \Dst(p_2, \eps_2)K(p_3)|i\mathcal{L}_{\rm eff}|D_{sJ}(2700)(p_1,\eps_1)\rangle = &
    -i\tilde{g}_{D_{sJ} \Dst K}\,
      \epsilon_{\mu \nu \alpha \beta} \eps_1^{\mu}\eps_2^{\nu} p_3^{\alpha}p_1^{\beta},
\end{split}
 \label{eq:DsJ_matrixelement}
\end{equation}
where the strong coupling constants are given by Ref.~\cite{Colangelo:2007ds},
%% DsJ form factors
\begin{equation}
\begin{split}
\tilde{g}_{D_{sJ}DK} & = 2\frac{\tilde{g}}{f_{\pi}}\sqrt{m_{D_{sJ}}m_D},\\
\tilde{g}_{D_{sJ}\Dst K} & = 2\frac{\tilde{g}}{f_{\pi}}\sqrt{\frac{m_{\Dst}}{m_{D_{sJ}}}},
\end{split}
 \label{eq:DsJ_gtilde}
\end{equation}
with $f_{\pi}=132\ \text{MeV}$.
Once the coupling constants and form factors are extracted,
one can insert
Eq.~\eqref{eq:DsJ_matrixelement} into Eq.~\eqref{eq:pole_derivation}
to obtained the contribution to form factors from the $D_{sJ}(2700)$ resonance.

From these matrix elements, Ref.~\cite{Colangelo:2007ds} predicted
the ratio of branching fractions
\begin{equation}
 \begin{split}
   r(\Dst K)
   = 0.91\pm0.04.
 \end{split}
 \label{eq:DsJ_br_ratio_theory}
\end{equation}
This ratio agrees with Eq.~\eqref{eq:DsJ_br_ratio_exp} very well.
The ratios of the branching fractions of the six main decay modes
are given by Table~\ref{table:DsJ_br}. The mixing angle between $\eta$ and $\eta'$ is taken
from Ref.~\cite{Feldmann:1998sh}.

%%%%%%%%%%%%%%%%%%%%%%%%%%%%%%%%%%%%%%%%%%%%%%%%%%%%%
%% Table I: DsJ decay mode branching ratios
\begingroup
%\squeezetable
\begin{table*}[t]
\begin{tabular}{p{3cm} p{1.5cm} p{1.5cm} p{1.5cm} p{1.5cm} p{1.5cm} p{1.5cm}}
\hline \hline
Mode(f)
& $D^0 K^+$ & $D^+ \Kbar^0$ & $D^{*0} K^{+}$ & $D^{*+} \Kbar^{0}$
& $D_s \eta$ & $\Dst_s \eta$  \\
\hline
$r(f)$ & 1.02 & 0.98 & 0.93 & 0.89 & 0.17 & 0.04 \\
\hline \hline
\end{tabular}
\caption{The ratio $r$ of the
branching fractions of six main decay modes of the $D_{sJ}(2700)^+$ resonance.}
\label{table:DsJ_br}
\end{table*}
\endgroup
%%%%%%%%%%%%%%%%%%%%%%%%%%%%%%%%%%%%%%%%%%%%%%%%%%

Assuming $D_{sJ}(2700)$ only decays to $D^{(*)}K$ and $D^{(*)}\eta^{(\prime)}$,
$\tilde{g}^2$ is proportional to the total width.
Thus, we have
\begin{equation}
 \tilde{g} = 0.28 \pm0.03,
 \label{eq:gtilde}
\end{equation}
where the uncertainty comes from the uncertainty of the total width.
Note that this value is slightly larger
than the one in Ref.~\cite{Colangelo:2007ds}
as the world-average of width [Eq.~\eqref{eq:DsJ_masswidth}] became larger.

Taking the measured mass, width and $\mathcal{B}(B \to \Dbar^{(*)}D_{sJ}(2700))\times
\mathcal{B}(D_{sJ}(2700) \to DK)$ (see Sec.~III.~B for details) as input,
the $D_{sJ}(2710)$ decay constant is extracted as
\begin{equation}
f_{D_{sJ}(2700)}  = 240\pm 31 \text{ MeV}.
\end{equation}
The decay constant can be compared to the previous estimations
$243\pm41~ \text{MeV}$ in Ref.~\cite{Colangelo:2007ds} and
$295\pm13~ \text{MeV}$ in Ref.~\cite{Wang:2009zz}.
Note that it is compatible to the decay constants of $D^{(*)}_s$,
which we use $260\pm13~ \text{MeV}$ in later calculation.

The $\Bbar_s \to D_{sJ}(2700)$ transition form
factors can be obtained by using a covariant light-front quark model~\cite{Cheng:2004ru}.
For the $2S$ wave function,~\footnote{
In the quark model with a simple harmonic like potential, the wave function
for a state with the quantum numbers $(n,l,m)$ is given by
$f_{nl}(\vec{p}^2/\beta^2)Y_{lm}(\hat{p})\text{exp}(-\vec{p}^2/2\beta)$
with $f_{10}(x)=1$ and $f_{20}(x)=\sqrt{\frac{3}{2}}(-1+\frac{2}{3}x)$.
We fit the Gaussian width $\beta$ to decay constant.}
its Gaussian width can be fixed
by the decay constant derived from Eq.~\eqref{eq:DsJ_partialbr_exp}.
It is then straightforward to obtain various $\bar B_s\to D_{sJ}$ form factors:
\begin{equation}
\begin{split}
V^{\Bbar_s D_{sJ}(2700)}(q^2) &
=\frac{0.25\pm0.03}{1-0.03\,q^2/m_{Bs}^2+0.38\,q^4/m_{Bs}^4},\\
A^{\Bbar_s D_{sJ}(2700)}_0(q^2) &
=\frac{0.24\pm0.02}{1+1.16\,q^2/m_{Bs}^2+2.16\,q^4/m_{Bs}^4},\\
A^{\Bbar_s D_{sJ}(2700)}_1(q^2) &
=\frac{0.17\pm0.02}{1+0.66\,q^2/m_{Bs}^2+0.54\,q^4/m_{Bs}^4},\\
A^{\Bbar_s D_{sJ}(2700)}_2(q^2) &
=\frac{0.007\pm0.001}{1+4.84\,q^2/m_{Bs}^2+5.08\,q^4/m_{Bs}^4}.\\
\end{split}
 \label{eq:DsJ_FF}
\end{equation}
These transition form factors are small comparing to the $D^{(*)}_s$
(collected in Appendix~\ref{basic_parameters}),
because of the poor overlap between wave functions of
ground state $B$ mesons and the radial excited $D_{sJ}(2700)$.

\subsection{Non-Resonance Contribution}

In general, there will be both resonant and non-resonant (NR) contributions to form
factors. In previous study of $\Bbar \to D^{(*)}K^-K^0$
decays~\cite{Chua:2002pi}, it is necessary to add NR contribution to
form factors to explain the experimental observations.
Therefore, we should include the NR effect in this work.
To produce the $D^{(*)}K^{(*)}$ pairs, at least one gluon must be
emitted to produce $q\bar{q}$ pairs.
The QCD counting rule~\cite{Chua:2002pi} provides an ansatz for the
asymptotic behavior of the non-resonant form factors,
which is
%% Non-resonance behavior of form factors
\begin{equation}
F(q^2)_{NR} \rightarrow \frac{x_F}{q^2}\left[\text{ln}\left(\frac{q^2}{\Lambda^2}\right)\right]^{-1},
 \label{eq:nonres_asymp}
\end{equation}
where $q^2$ is the invariant mass of $D^{(*)}K^{(*)}$ and
$\Lambda=0.5\text{ GeV}$ is the QCD scale.

Together with the pole contribution provided in Appendix~\ref{full_pole},
the complete form factors are modeled by the pole and NR contribution,
%% Total Form Factors
\begin{equation}
F(q^2) = F(q^2)|_{\rm pole} + \frac{x_F}{q^2}\left[\text{ln}\left(\frac{q^2}{\Lambda^2}\right)\right]^{-1},
 \label{eq:pole_and_non}
\end{equation}
where the asymptotic form of NR contribution is adopted for simplicity.
As more data is available in the future, one could replace this simple form
with a more sophisticated one
to fit the data, as in Ref.~\cite{Chua:2002pi}.

%%%%%%%%%%%%%%%%%%%%%%%%%%%%%%%%%%%%%%%%%%%%%%%%%%%%%%%%%
\section{Results}
\subsection{Two-body $\mathcal{D}^{(*)}_s \mathcal{\bar{D}}^{(*)}_s$ Decays
and the Width Difference: An Update}
\label{sec:twobody_results}
%%%%%%%%%%%%%%%%%%%%%%%%%%%%%%%%%%%%%%%%%%%%%%%%%%%%%%%%%

We first update the branching fractions of two-body $\bar B_s\to D^{(*)}_s \Dbar^{(*)}_s$ decays,
which contribute to $\DGamma_s$.
The necessary parameters are given in Appendix~\ref{basic_parameters}.
Our results are listed in Table~\ref{table:twobody_results},
where experimental results and previous theoretical results from
Ref.~\cite{Aleksan:1993qp} are listed for comparison.
Since SU(3)-related modes in $B_{u,d}$ systems are usually more precisely known than the $B_s$ system,
we also list them in parentheses for comparison.
For example, data for $\Br(\Bbar_u \to D_u \Dbar_s)$,
which is approximately the same as $\Br(\Bbar_s \to D_s \Dbar_s)$
under SU(3) limit, is listed.
Note that two uncertainties are given in our results:
The first uncertainty is obtained by varying decay constants and form factors by 5\%,
while the second comes from the estimated $10\%$ uncertainty in $a_1$.

%%%%%%%%%%%%%%%%%%%%%%%%%%%%%%%%%%%%%%%%%%%%%%%%
%% Table II: Result of two-body states
\begingroup
%\squeezetable
\begin{table*}[t]
\begin{tabular}{c p{3cm} p{2.4cm} p{2.4cm} p{2.4cm} p{2.4cm}}
\hline \hline
Mode(f)  & $\mathcal{B}(\Bbar_{s,(u)} \to f)$ (\%) \newline{data}
& $\mathcal{B}(\Bbar_s \to f)$ (\%) \newline{this work}
& $\mathcal{B}(\Bbar_s \to f)$ (\%) \newline{Ref.~\cite{Aleksan:1993qp}}
& $\DGamma_f / \Gamma_s$ (\%) \newline{this work}
& $\DGamma_f/  \Gamma_s$ (\%) \newline{Ref.~\cite{Aleksan:1993qp}}\\
\hline
$D_s \Dbar_s$
& \err{1.04}{0.35}~\footnotemark[1]
\newline{(\err{1.00}{0.17})~\footnotemark[1]}
& \errr{1.4}{0.3}{0.3} & 1.6 & \errr{2.7}{0.6}{0.6} & 3.1 \\
$D_s^* \Dbar_s$+$D_s \Dbar_s^*$
& \err{2.75}{1.08}~\footnotemark[2]
\newline{($1.58\pm0.33$)~\footnotemark[1]}
& \errr{1.8}{0.4}{0.4} & 2.2 & \errr{3.6}{0.8}{0.8} & 4.4  \\
$D_s^* \Dbar_s^*$
& \err{3.08}{1.49}~\footnotemark[2]
\newline{(\err{1.71}{0.24})~\footnotemark[1]}
& \errr{2.3}{0.5}{0.5} & 3.6 & \errr{3.8}{0.8}{0.8} & 6.9  \\
\hline
$D_s^{(*)} \Dbar_s^{(*)}$
& \err{4.9}{1.4}~\footnotemark[3]
\newline{\err{6.9}{2.3}~\footnotemark[2]}
\newline{\err{4.0}{1.5}~\footnotemark[1]}
\newline{(\err{4.29}{0.74})~\footnotemark[1]}
& \errr{5.5}{1.2}{1.1}   & 7.4 & \errr{10.2}{2.2}{2.1} & 14.4 \\
%\hline
%$D_s^{(*)} \Dbar_s^{**}$, $D_s^{**} \Dbar_s^{(*)}$,$D_s^{**} \Dbar_s^{**}$
%& N/A               & \errr{2.6}{0.7}{0.5} & N/A & \errr{0.2}{0.3}{0.04} &N/A \\
\hline \hline
\end{tabular}
\caption{
The branching fractions of $\Bbar_s \to D^{(*)}_s\Dbar^{(*)}_s$ decays
and their contribution to the width difference.
The results can be compared with data in Refs.~\cite{PDG, Esen:2010jq, HFAG:2010}.
The data for $B^-$ system in Ref.~\cite{PDG},
which are related to $B_s$ under SU(3) symmetry,
are shown in parentheses (see text for detail).
The theoretical result of Ref.~\cite{Aleksan:1993qp}
is also presented for comparison.
}
\footnotetext[1]{Data taken from Ref.~\cite{PDG}.}
\footnotetext[2]{Data taken from Ref.~\cite{Esen:2010jq}.}
\footnotetext[3]{Data taken from Ref.~\cite{HFAG:2010}.}
\label{table:twobody_results}
\end{table*}
\endgroup
%%%%%%%%%%%%%%%%%%%%%%%%%%%%%%%%%%%%%%%%%%%%%%%%%%

The branching fractions of $D_s^{(*)} \Dbar_s^{(*)}$ modes are all
of percent level.
In general, our result is smaller than the result
in Ref.~\cite{Aleksan:1993qp}.
These branching fractions can be compared with experimental data in both
$B_s$ and $B^-$ system.
One can see that our results agree with experiment within uncertainties.
The direct measurement of $\Bbar_s \to D_s^{(*)} \Dbar_s^{(*)}$
exclusive decays was recently reported by Belle~\cite{Esen:2010jq}.~\footnote
{Note that this measurement does not tag the flavor of the $B_s$ meson.
Although there should be a corresponding
correction to the order of $\DGamma_s/\Gamma_s$~\cite{Dunietz:2000cr},
it is smaller than the theoretical errors and omitted from the table.}
While the observed branching fraction of $D_s \Dbar_s$ mode
$(1.0 \pm 0.4) \%$ is close to our result,
other modes are more aligned with the calculation in Ref.~\cite{Aleksan:1993qp}.
But the world average of the inclusive branching fraction
$\Br(\Bbar_s \to D_s^{(*)} \Dbar_s^{(*)})$ ~\cite{HFAG:2010,PDG} and
the rates of SU(3) related modes are closer to our results.

%%%%%%%%%%%%%%%%%%%%%%%%%%%%%%%%%%%%%%%%%%%%%%%%%%%
%% Table II: Result of two-body states with Ds0, Ds1, Ds1*
\begingroup
%\squeezetable
\begin{table*}[t]
\begin{tabular}{c p{3.2cm} p{3.2cm} p{3.2cm}}
\hline \hline
Mode(f)
& $\mathcal{B}(\Bbar_s \to f)$ (\%)
& $\mathcal{B}(B_s \to f)$ (\%)
& $\DGamma_f / \Gamma_s$ (\%) \\
\hline
$D_s \Dbar_{s0}^*(2317)$
& \errr{0.10}{0.02}{0.02} \newline{($0.073^{+0.022}_{-0.017}$)~\footnotemark[1] }
	& \errr{0.15}{0.03}{0.03} & \errr{-0.24}{0.05}{0.05}\\
$\Dst_s \Dbar_{s0}^*(2317)$
& \errr{0.05}{0.01}{0.01} \newline{($0.09\pm0.07$)~\footnotemark[1] }
	& \errr{0.12}{0.03}{0.03} & \errr{-0.15}{0.03}{0.03} \\
$D_s \Dbar_{s1}(2460)$
& \errr{0.24}{0.05}{0.05} \newline{($0.31^{+0.10}_{-0.09}$)}
	& \errr{0.04}{0.01}{0.01} & \errr{-0.18}{0.04}{0.04}\\
$\Dst_s \Dbar_{s1}(2460)$
& \errr{0.81}{0.17}{0.17} \newline{($1.20\pm0.30$)}
	& \errr{0.06}{0.01}{0.01} & \errr{+0.16}{0.03}{0.03} \\
$D_s \Dbar_{s1}(2536)$
& \errr{0.02}{0.01}{0.01} \newline{($0.022\pm0.007$)~\footnotemark[2]}
	& \errr{0.38}{0.08}{0.08} & \errr{+0.19}{0.04}{0.04} \\
$\Dst_s \Dbar_{s1}(2536)$
& \errr{0.09}{0.02}{0.02} \newline{($0.055\pm0.0016$)~\footnotemark[2]}
	& \errr{0.38}{0.08}{0.08}& \errr{+0.34}{0.07}{0.07} \\
$D_{s0}^*(2317) \Dbar_{s1}(2460)$
& \errr{0.024}{0.005}{0.005} & \errr{0.002}{0.001}{0.001}
        & \errr{+0.013}{0.003}{0.003}\\
$D_{s0}^*(2317) \Dbar_{s1}(2536)$
&  \errr{0.002}{0.001}{0.001} & \errr{0.017}{0.004}{0.004}
        & \errr{-0.012}{0.003}{0.003} \\
$D_{s1}(2460) \Dbar_{s1}(2536)$
& \errr{0.001}{0.001}{0.001} & \errr{0.077}{0.017}{0.016}
        & \errr{+0.000}{0.000}{0.000}\\
$D_{s0}^*(2317) \Dbar_{s0}^*(2317)$
& \multicolumn{2}{c}{\errr{0.009}{0.002}{0.002} }
        & \errr{+0.018}{0.004}{0.004} \\
$D_{s1}(2460) \Dbar_{s1}(2460)$
& \multicolumn{2}{c}{\errr{0.014}{0.003}{0.003} }
        & \errr{-0.010}{0.002}{0.002} \\
$D_{s1}(2536) \Dbar_{s1}(2536)$
& \multicolumn{2}{c}{\errr{0.007}{0.002}{0.001} }
        & \errr{+0.008}{0.002}{0.002} \\
\hline
Total
& \multicolumn{2}{c}{\errr{2.57}{0.55}{0.54} }
& \errr{0.24}{0.27}{0.05}~\footnotemark[3] \\
\hline \hline
\end{tabular}
\caption{The branching fractions and width difference of
$\Bbar_s$ and $\Bbar_s$ decays to two-body $D^{(*,**)}_s\Dbar_s^{**}$,
where $D^{**}_s$ is $D_{s0}^*(2317)$, $D_{s1}(2460)$, or $D_{s1}(2536)$.
We show data of SU(3) related modes in $\Bbar_{u}$ system ~\cite{PDG} in parentheses for comparisons.
}
\footnotetext[1]{$\mathcal{B}(B^-\to D^{(*)0}\bar D_{s0}(2317))\times\mathcal{B}(\bar D_{s0}(2317)\to \bar D_s\pi^-)$.}
\footnotetext[2]{$\mathcal{B}(B^-\to D^{(*)0}\bar D_{s1}(2536))\times\mathcal{B}(\bar D_{s1}(2536)\to \bar D^*K^-)$.}
\footnotetext[3]{The contribution from $CP$ conjugate modes
$\bar{f}$ is included.}
\label{table:twobody_Dstst}
\end{table*}
\endgroup
%%%%%%%%%%%%%%%%%%%%%%%%%%%%%%%%%%%%%%%%%%%%%

The total $\DGamma_f / \Gamma_s$ induced by $D_s^{(*)} \Dbar_s^{(*)}$ modes
is $10.2\pm2.2\pm2.1\%$.
This value is smaller than the
previous long-distance calculation~\cite{Aleksan:1993qp} also shown in this table.
In addition, the total $\DGamma_f / \Gamma_s$ does not reach
the short-distance central value in Eq.~\eqref{eq:DGamma_sd}.
One also observes that $\DGamma_s(D_s^{(*)} \Dbar_s^{(*)}) / \Gamma_s$
is approximately two times the total branching fractions.
The relation $|\DGamma_s(f) / \Gamma_s|\leq  2\sqrt{\Br(\bar B_s\to f)\Br(B_s\to f)}$,
which corresponds to the maxima in Eq.~\eqref{eq:DGamma_limit_3},
saturates only when the mode(s) $f$ is purely $CP$-even,
such as the $D_s \Dbar_s$ mode.
The nearly maximal $\DGamma_f$ reflects that $D_s^{(*)}$ are very efficient
in mediating the width difference.

Several new $c\bar s$ resonances are found in $B$ decays. They may also contribute to $\Delta\Gamma_s$.
We calculate the contribution by the
two-body modes with $D_{s0}^*(2317)$, $D_{s1}(2460)$, and $D_{s1}(2536)$.
Results are shown in Table~\ref{table:twobody_Dstst}.
There are additional 21 modes when these higher $D_s^{**}$ resonances are
included. Note that not all modes are shown explicitly in the Table.
Since $\CP$ is conserved in this work,
$\mathcal{B}(\Bbar _s\to f) = \mathcal{B}(B_s \to \bar{f})$ and
$\DGamma_f = \DGamma_{\bar{f}}$.
For modes which are not $\CP$ eigenstates, the contributions from their $\CP$ conjugates
are also known and should be added to $\DGamma_s/\Gamma_s$.
The total branching fraction of these additional modes is comparable to
the sum of $\mathcal{B}(D^{(*)}_s\bar D^{(*)}_s)$.
However, the corresponding contribution to the width difference turns out to be tiny.
After considering all of these two-body modes,
the total $\DGamma_f / \Gamma_s$ only increase slightly from
$10.2\pm2.2\pm2.1\%$ to $10.4\pm2.5\pm2.2\%$.
There are two reasons for such a tiny contribution.
First, the sign of $\DGamma_f$ are fluctuating among these modes,
leading to cancellations in the total sum.
In addition, the ``mismatch'' effect is serious.
For instance, the $\Bbar_s \to D_s^* \Dbar_{s1}(2460)$ mode has a
non-negligible branching fraction $0.81\%$,
but the branching fraction of
$B_s \to D_s^* \Dbar_{s1}(2460)$ is only $0.06\%$.
In fact, the smallness of contributions  in the heavy quark limit from $p$-wave resonances
was expected~\cite{Aleksan:1993qp}, and is confirmed in a realistic calculation given here.

The sizable branching fraction
$\mathcal{B}(\bar B \to D^{(*)}\bar D_{sJ}(2700))\times
\mathcal{B}(\bar D_{sJ}(2700) \to \bar D\bar K)$
indicates that the $\bar D_{sJ}(2700)$ resonance may be important for $\DGammas$.
Since $\bar D_{sJ}(2700)$ has a broad width,
it is expected to interfere with
the continuum of $\Bbar_s \to D_s\bar D^{(*)}\bar K$ produced by $\bar D^{(*)}_s$ poles (see Fig.~\ref{fig:pole_diagram} and the next subsection).
For completeness, it is better to calculate the contribution of $\bar D_{sJ}(2700)$ to
$\Delta \Gamma_s$ in three-body modes,
including the on-shell and off-shell parts.
However, the two-body calculation is simple and straightforward.
It is, therefore, helpful to see the contribution of $\bar D_s^{(*,**)} D_{sJ}(2700)$
to $\Delta \Gamma_s$ first.

Using the parameters calculated
in Eq.~\eqref{eq:DsJ_FF}, the contributions from  two-body modes including $\bar D_{sJ}(2700)$ is
shown in Table~\ref{table:DsJ_twobody}.
Several things ought to be noted:
(a) The branching fractions of modes with current-produced $\Dbar_{sJ}(2700)$
(the $\mathcal{B}(\Bbar_s \to f)$ column of Table.~\ref{table:DsJ_twobody})
are comparable to those of the $D^{(*)}_s\Dbar^{(*)}_s$ modes.
The two-body decays with $\bar D_{sJ}(2700)$ seem to be suppressed seriously
by phase space at first glance.
Nevertheless, this may not be true since the factorized amplitude
$\langle \mathcal{D}^{*}_s |(V-A)_{\mu}|0\rangle$
[see Eq.~(\ref{eq:D_decayconst})] for current-produced meson
is enhanced by mass, and the decay constant of $\bar D_{sJ}(2700)$ is unsuppressed.
(b) For the mode $\Bbar_s \to \Dst_s \Dbar_{sJ}(2700)$, there are several enhancement and suppression factors,
when replacing $D^*_s$ with $D_{sJ}(2700)$.
First, its amplitude is dominated by $s$-wave and is free from additional momentum suppression. In addition,
it is enhanced through the above mentioned factorized amplitude and suppressed by phase space.
The branching fraction of $\Bbar_s \to \Dst_s \Dbar_{sJ}(2700)$
turns out to decrease $\sim 10\%$ compared with $\Bbar_s \to \Dst_s \Dbar^*_s$.
On the contrary, the decay $\Bbar_s \to D_s \Dbar_{sJ}(2700)$ is $p$-wave.
Its amplitude and thus branching fraction drops more than 50\%
when compared to $\Bbar_s \to D_s \Dbar^*_s$.
The two different trends lead to a large ratio
$\Br(\Bbar_s \to \Dst_s \Dbar_{sJ}(2700)) /\Br(\Bbar_s \to D_s \Dbar_{sJ}(2700))
\approx 5$.
(c) The branching fractions of modes
in which $\Dbar_{sJ}(2700)$ contains the spectator quark
(the $\mathcal{B}(B_s \to f)$ column) are very small.
%The above mentioned enhancement does not happens
%in modes in which $D_{sJ}(2700)$ contains the spectator quark.
The branching fractions are suppressed not only by phase space,
but also by the small transition form factors shown in Eq.~\eqref{eq:DsJ_FF}.

%%%%%%%%%%%%%%%%%%%%%%%%%%%%%%%%%%%%%%%%%%%%%%%%%%%%
%% Table IV: Result of two-body DsJ states
\begingroup
%\squeezetable
\begin{table*}[t]
\begin{tabular}{c p{2.8cm} p{2.8cm} p{2.8cm}}
\hline \hline
Mode(f)
& $\mathcal{B}(\Bbar_s \to f)$ (\%)
& $\mathcal{B}(B_s \to f)$ (\%)
& $\DGamma_f / \Gamma_s$ (\%) \\
\hline
$D_s \Dbar_{sJ}(2700)$
& \errr{0.44}{0.18}{0.09} & \errr{0.02}{0.01}{0.01} & \errr{0.21}{0.08}{0.04}\\
$\Dst_s \Dbar_{sJ}(2700)$
& \errr{2.0}{0.8}{0.4} & \errr{0.08}{0.03}{0.02} & \errr{0.73}{0.27}{0.15} \\
\hline
$D^{(*)}_s \Dbar_{sJ}(2700)$
& \errr{2.5}{1.0}{0.5} & \errr{0.11}{0.03}{0.02}
& \errr{1.9}{0.7}{0.4}~\footnotemark[1]\\
\hline
$D^{**}_s \Dbar_{sJ}(2700)$
& \errr{0.14}{0.08}{0.03} & \errr{0.02}{0.07}{0.01}
& \errr{0.08}{0.03}{0.02}~\footnotemark[1] \\
\hline \hline
\end{tabular}
\caption{The branching fractions and width difference of two-body
$\Bbar_s$ and $B_s$ decays to  $D^{(*,**)}_s\Dbar_{sJ}(2700)$, where
$D^{**}_s$ stands for $D_{s0}^*(2317)$, $D_{s1}(2460)$, or $D_{s1}(2536)$.
}
\footnotetext[1]{The contribution from $CP$ conjugate modes
$\bar{f}$ is included.}
\label{table:DsJ_twobody}
\end{table*}
\endgroup
%%%%%%%%%%%%%%%%%%%%%%%%%%%%%%%%%%%%%%%%%%%%%%%%%%%%%%

The $\DGamma_s$ from $D^{(*)}_s \Dbar_{sJ}(2700)$ is $1.9\pm0.7\pm0.4\%$.
As the upper bound in Eq.~\eqref{eq:DGamma_limit_3} implies,
the $\DGamma_s/\Gamma_s$ of $\Dbar_{sJ}(2700)$ is limited by
the imbalance between the modes
in which $\bar D_{sJ}(2700)$ produced via current or with spectator.
Nevertheless, the contribution form $D_{sJ}(2700)$ is larger than those from $D_s^{**}$
and should not be neglected.
We remark that, as we shall see in the three-body case,
the transition amplitudes from $D^{(*)}_s$ poles can interfere constructively
with the current-produced $D_{sJ}$ pole and overcome the above mentioned suppression,
leading to sizable contribution to $\Delta\Gamma_s$.

%%%%%%%%%%%%%%%%%%%%%%%%%%%%%%%%%%%%%%%%%%%%%%%%%%%%%%%%%
\subsection{Three-body $D^{(*)}_s \Dbar^{(*)} \Kbar^{(*)}$ Decays and
Contributions to the Width Difference}
\label{sec:threebody_results}
%%%%%%%%%%%%%%%%%%%%%%%%%%%%%%%%%%%%%%%%%%%%%%%%%%%%%%%%%

We now turn to the three-body case.
We shall first compare our results with the measured branching fractions
in $B_{u,d}$ system, starting from pole model and
including NR effect, if necessary.
After demonstrating that our calculation is consistent with data,
we proceed to calculate the width difference in the $B_s$ system.

\subsubsection{Current-Produced Branching Fractions in $B_{u,d}$ systems}

Only current-produced modes with $\bar K$ have been measured in $\bar B_{u,d}$ systems.
There is no measurement for the rest of the modes,
including current-produced
$\bar \Kst$, and all the transition modes.
A summary of current data and our results is presented in
Table~\ref{table:threebody_exp_results}.
We separate the results of BaBar and Belle for comparison.
Note that in $\Bbar_{u,d}$ systems,
some $D^{(*)}\Dbar^{(*)}\Kbar^{(*)}$ modes contain both color allowed and
color-suppressed diagrams,
where the latter is  expected to be sub-leading and is neglected in this work.
We labelled these modes in the remarks of the table,
and also add approximation sign in front of our results.
Note that in the calculation of $\DGamma_s$ in $\bar B_s$ system,
color-suppressed diagrams only appear in modes with $\eta^{(\prime)}$ and do not affect
$\DDKgeneral$ modes.

%%%%%%%%%%%%%%%%%%%%%%%%%%%%%%%%%%%%%%%%%%%%%%%%%%%%
%% TableV: Exp. Result Comparison
\begingroup
%\squeezetable
\begin{table*}[t]
\begin{tabular}{p{4cm}  p{2.3cm} p{2.3cm} p{3.2cm} p{2.8cm} p{3.0cm}}
\hline \hline
Measurement & BaBar Data(\%) & Belle Data(\%) &
\multicolumn{2}{l}{Our Results (\%)} & Remarks \\
  &  &  & Scenario \Rmnum{1} (\Rmnum{1}$^\prime$) & Scenario \Rmnum{2} & \\
  &  &  & Pole model with $D_{sJ}$ & Pole model+NR & \\
  &  &  & (without $D_{sJ}$) &   & \\
\hline
\multicolumn{6}{c}
{Category 1: current-produced $\Dbar\Kbar$  with $\Bbar \to D$ transition} \\
$\Br(\Bbar_u \to D_u \Dbar_{sJ}(2700)^-) \times$
  \newline{$\quad \Br(\Dbar_{sJ}(2700)^- \to \Dbar^0K^-)$}
&  N/A
&  $0.113^{+ 0.026}_{-0.040}$\footnotemark[2]
&  \errr{0.12}{0.08}{0.03} \newline{(0)} & \errr{0.12}{0.08}{0.03}
&  Input for Scenario \Rmnum{1}\\

$\Br(\Bbar_u \to D_u \Dbar^0K^-)$
&  $0.131\pm 0.014$\footnotemark[1]
&  $0.222\pm 0.033$\footnotemark[2]
&  $\sim0.23$ \newline{($\sim0.07$)} & $\sim0.11$
&  Color-suppressed diagram neglected\\

$\Br(\Bbar_d \to D_d \Dbar^0K^-)$
&  $0.107\pm 0.011$\footnotemark[1] &  N/A
& \errr{0.22}{0.14}{0.05} \newline{(\errr{0.06}{0.03}{0.01})}
& $0.10^{+0.23}_{-0.02}\pm{0.02}$
& Input for Scenario \Rmnum{2}\\

$\Br(\Bbar_d \to D_d \Dbar_{sJ}(2700)^-) \times$
  \newline{$\quad \Br(\Dbar_{sJ}(2700)^- \to \Dbar^0K^-)$}
&  N/A &  N/A
& \errr{0.11}{0.07}{0.02} \newline{(0)} & \errr{0.11}{0.07}{0.02}  &  \\

\hline
\multicolumn{6}{c}
{Category 2: current-produced $\Dbar\Kbar$ with $\Bbar \to D^*$ transition} \\
$\Br(\Bbar_d \to D^*_d \Dbar^0K^-)$
&  $0.247\pm 0.021$\footnotemark[1] &  N/A
&  \errr{0.67}{0.45}{0.14} \newline{(\errr{0.07}{0.03}{0.01})}
& ${0.32}^{+0.75}_{-0.13}\pm{0.07}$
& Input for Scenario \Rmnum{2}\\

$\Br(\Bbar_d \to D^*_d \Dbar_{sJ}(2700)^-) \times$
  \newline{$\quad \Br(\Dbar_{sJ}(2700)^- \to \Dbar^0K^-)$}
&  N/A &  N/A & \errr{0.50}{0.33}{0.11}\newline{(0)}
& \errr{0.50}{0.33}{0.11}  &  \\

\hline
\multicolumn{6}{c}
{Category 3: current-produced $\Dbar^*\Kbar$ with $\Bbar \to D$ transition} \\
$\Br(\Bbar_d \to D_d \Dbar^{*0}K^-)$
&  $0.346\pm 0.041$\footnotemark[1] &  N/A
&  \errr{0.35}{0.21}{0.07} \newline{(\errr{0.20}{0.10}{0.04})}
&  \errr{0.35}{0.21}{0.07}\footnotemark[5] & \\

$\Br(\Bbar_d \to D_d \Dbar_{sJ}(2700)^-) \times$
  \newline{$\quad \Br(\Dbar_{sJ}(2700)^- \to \Dbar^{*0}K^-)$}
&  N/A &  N/A
& \errr{0.11}{0.07}{0.02} \newline{(0)}
& \errr{0.11}{0.07}{0.02}\footnotemark[5]  &  \\

\hline
\multicolumn{6}{c}
{Category 4: current-produced $\Dbar^*\Kbar$ with $\Bbar \to D^*$ transition} \\
$\Br(\Bbar_d \to D^*_d \Dbar^{*0}K^-)$
&  $1.060\pm 0.092$\footnotemark[1] &  N/A
&  \errr{0.94}{0.62}{0.20} \newline{(\errr{0.15}{0.08}{0.03})}
&  \errr{0.94}{0.62}{0.20} \footnotemark[5] &  \\

$\Br(\Bbar_d \to D^*_d \Dbar_{sJ}(2700)^-) \times$
  \newline{$\quad \Br(\Dbar_{sJ}(2700)^- \to D^{*0}K^-)$}
&  N/A &  N/A
& \errr{0.52}{0.33}{0.11}\newline{(0)}
& \errr{0.52}{0.33}{0.11}\footnotemark[5] & \\

$\Br(\Bbar_d \to D^*_d \Dbar^{*+}\Kbar^0)$
&  $0.826\pm 0.080$\footnotemark[1]
&  N/A
&  $\sim0.91$\newline{($\sim0.15$)} & $\sim0.91$\footnotemark[5]
&  Color-suppressed diagram neglected\\

$\Br(\Bbar_d \to D^*_d \Dbar^{*+}K^0_S)$
&  $0.44\pm 0.08$\footnotemark[3]
&  $0.34\pm 0.08$\footnotemark[3]
&  $\sim0.46$\newline{($\sim0.07$)} & $\sim0.46$\footnotemark[5]
& Color-suppressed diagram neglected\\

\hline \hline
\end{tabular}
\caption{Comparison between experimental results from BaBar and
Belle collaborations and our results in Scenario \Rmnum{1}, \Rmnum{2},
and \Rmnum{1}$^\prime$. See text for detailed definition.}
\footnotetext[1]
             {Ref.~\cite{delAmoSanchez:2010pg}.}
\footnotetext[2]
             {Ref.~\cite{Brodzicka:2007aa}.}
\footnotetext[3]
             {Ref.~\cite{Aubert:2006fh}.}
\footnotetext[4]
             {Ref.~\cite{Aubert:2007rva}.}
\footnotetext[5]{In Scenario \Rmnum{2},
the results of modes in Category 3,4 are the same as Scenario \Rmnum{1}.}
\label{table:threebody_exp_results}
\end{table*}
\endgroup
%%%%%%%%%%%%%%%%%%%%%%%%%%%%%%%%%%%%%%%%%%%%%%%%%%

According to whether $D$ or $D^*$, there are four types of $D^{(*)} \Dbar^{(*)} K$ modes, which are classified into four categories as shown in Table~\ref{table:threebody_exp_results}.
Modes in each category have similar branching fractions
because of SU(2) symmetry.
The measured branching fractions increase from Category 1 ($\sim 0.1\%$)
to Category 4 ($\sim 1\%$).
One can find tension
in measurements of $\Bbar_u \to D_u \Dbar^0K^-$.
A large $\bar D_{sJ}(2700)$ contribution has been observed in $\Bbar_u \to D_u \Dbar^0K^-$
by Belle only~\cite{Brodzicka:2007aa}, but in $2.2\sigma$ disagreement with
BaBar~\cite{delAmoSanchez:2010pg}.
The tension in data becomes more severe if one compares the $\bar D_{sJ}(2700)$
contribution to the total branching fraction of $\Bbar_u \to D_u \Dbar^0K^-$.
In the case of Belle,
the contribution from $\bar D_{sJ}(2700)$
is about half the total branching fraction.
However, it is approximately equal to the total branching fraction for BaBar.
As we show, the inconsistency makes it difficult to explain all data with
a simple pole model.

The results of our calculation in different scenarios
are compared with experiments in Table~\ref{table:threebody_exp_results}.
%Scenario \Rmnum{1} and \Rmnum{1}$^\prime$ are pole model results, while Scenario \Rmnum{2} includes NR contributions.
In Scenario \Rmnum{1}, $D_s^{(*)}$ and $D_{sJ}$ poles are used,
while in Scenario \Rmnum{1}$^\prime$,
only $D^{(*)}_s$ poles are considered,
with results shown in parentheses for comparison.
In Scenario \Rmnum{2}, NR contributions
in $\Dbar\Kbar$ time-like form factors are included to demonstrate that the inconsistency
with experiments in Scenario \Rmnum{1} can be resolved.
Note that no NR contribution is introduced for modes in Category 3 and 4
as the pole model results (Scenario \Rmnum{1}) already agree with data.
Furthermore, as there is no measurements on transition modes and modes with $\Kbar^*$,
no NR contribution is applied to these modes.
The two uncertainties of our results are obtained by the same method as in two-body case,
but with additional uncertainties from strong couplings included in the first errors.

Despite the disagreement between data,
we first attempt to explain all measurements
only with a pole model (Scenario \Rmnum{1}).
The corresponding diagrams can be found in the left portion of Fig.~\ref{fig:pole_diagram}
with the appropriate spectator quark.
In the calculation, we first fix the decay constant of $D_{sJ}(2700)$ from
the contribution of $\bar D_{sJ}(2700)$ in $\Bbar_u \to D_u \Dbar^0K^-$ decay.
The value of this decay constant is shown earlier in Eq.~\eqref{eq:DsJ_FF},
and the value agrees with those obtained in other studies (see Section~II.~D).
The total branching fraction of $\Bbar_u \to D_u \Dbar^0K^-$
is consistent with Belle's measurement, and inevitably less consistent with the BaBar result
and the SU(2)-related mode $\Bbar_d \to D_d \Dbar^0K^-$.
Unfortunately, there is no measurement on $\Bbar_d \to D_d \Dbar^0K^-$ rate from Belle yet.
For Category 2,
the total branching fraction $\Bbar_d \to D^*_d \Dbar^0K^-$ is about
2.5 times larger than the BaBar result as in Category~1.
Again, there is no measurement from Belle.
More data analysis is called for.
Nevertheless,
it is interesting to see that our predicted results on branching fractions in
Categories 3 and 4 agree well with data.

To explain the total branching fractions in Scenario \Rmnum{1},
we must start from the $\bar D_{sJ}(2700)$ contribution,
which has on-shell as well as off-shell parts.
Roughly speaking the $\bar D_{sJ}(2700)$ contribution can be understood by using the narrow width approximation.
The contribution in Category 1 (2) is almost the same as in Category 3 (4).
This is expected since the two categories are different from each other
only in $\bar D_{sJ}(2700)\to \bar D^*\bar K$, $\bar D\bar K$ parts,
which have nearly the same branching fractions [see Eq.~\eqref{eq:DsJ_br_ratio_theory}].
The contribution in Category 2 is about five times larger than
in Category~1, where the $\Bbar \to D^*$ transition is replaced with $\Bbar \to D$.
This factor already appeared in the two-body branching fractions of $\Bbar_s\to D^{(*)}_s \bar D_{sJ}$
modes shown in Table~\ref{table:DsJ_twobody}.
However, a closer look reveals that the precise $D_{sJ}(2700)$ contribution
should be obtained by integrating the full three-body phase space,
as the width of $D_{sJ}(2700)$ is of the order of $0.1\,\text{GeV}$,
which is not narrow enough compared with the three-body phase space.
(For instance, the decay $\Bbar_s \to D^{*}_s \Dbar_{sJ}(2700)$ with
$\Dbar_{sJ}(2700) \to \Dbar^{*}\Kbar$,
the invariant mass of $\Dbar^{*}\Kbar$ ranges
roughly from $2.5\,\text{GeV}$ to $3.3\,\text{GeV}$.
The Breit-Wigner function for $D_{sJ}(2700)$,
with a peak at $2.7\,\text{GeV}$,
cannot be approximated as a delta function since its peak is less than
$2$ times of width above the lower limit of
the invariant mass of $\Dbar^{(*)}\Kbar$.)
The numerical results usually show a $10\%$ overestimation
by narrow width approximation.
In addition, the $\bar D_{sJ}(2700)$ contribution
in $\Bbar_d \to D^*_d \Dbar^{*0}K^-$ is slightly greater than
$\Bbar_d \to D^*_d \Dbar^{0}K^-$, where the ratio in
Eq.~\eqref{eq:DsJ_br_ratio_theory} is the other way around.
This is due to the contribution from the off-shell part.
The off-shell contribution in high momentum region
favors $\bar D_{sJ}(2700) \to \bar D^*\bar K$ over
$\bar D_{sJ}(2700) \to \bar D\bar K$,
as one can see from the strong interaction matrix elements
in Eq.~\eqref{eq:DsJ_matrixelement}. The former coupling
is quadratic in momentum,
while the latter is only linear.
The numerical results show that the off-shell effect is about $10\%$.
This correction also echos our assertion that the contribution of
$D_{sJ}(2700)$ should be treated in a three-body picture.

The effect of off-shell $\bar D^{(*)}_s$ poles can be read from Scenario \Rmnum{1}$^\prime$
shown in parenthesis.
For the first two categories, only $\bar D_s^*$ pole contributes,
while for the latter two categories,
containing the current generated $\bar D^*\bar K$,
the $\bar D_s$ pole starts to contribute as well.
This explains why modes in Category~3 and 4 have larger branching fractions
in Scenario~\Rmnum{1}$^\prime$.
It is interesting to note that all branching fractions
in Scenario~\Rmnum{1}$^\prime$ are deficient in explain experimental results.
The $D_{sJ}(2700)$ resonance provides an important source
for the non-negligible three-body branching fractions
of current-produced modes.
Comparing with Scenario \Rmnum{1}, one finds the interference between
$D_{sJ}(2700)$ and $D^{(*)}_s$ poles are not negligible.
For example, in the $\bar B^0\to D^{*+}\bar D^{*0} K^-$  decay rate (see Category 4 in Table~\ref{table:threebody_exp_results}), the $\bar D^{(*)}_s$ and $\bar D_{sJ}$ contributions are $\sim0.15\%$ and $\sim0.52\%$, respectively, while the total predicted rate is
$\sim0.94\%$, which implies a fairly effective constructive interference between these poles.
If the $D_{sJ}$ width were narrow,
we would expect the interference effect to be negligible and it would be enough to consider a real $D_{sJ}(2700)$ in two-body final states.

After the above discussion, one can now understand the total branching fractions in Scenario \Rmnum{1}
by combining contributions of three different poles (see the left portion of Fig.~\ref{fig:pole_diagram}).
The contribution of $\bar D_{sJ}(2700)$ dominates over $\bar D^{(*)}_s$.
To first order,
Category 2 ($D^*\bar D\bar K$) and 4 ($D^*\bar D^*\bar K$)
have the same branching fractions from $\bar D_{sJ}(2700)$ and
are larger than Category 1 ($D\bar D \bar K$) and 3 ($D\bar D^*\bar K$).
$\bar D^{(*)}_s$ poles further split the two categories that have almost the same
$\bar D_{sJ}(2700)$ contribution.
Consequently, modes in Category 4 ($D^*\bar D^*\bar K$) have larger total branching fractions
than Category 2 ($D^*\bar D\bar K$), and similarly for Category 3 ($D\bar D^*\bar K$) and 1 ($D\bar D \bar K$).
The three different poles form the hierarchy of
total branching fractions of the four categories in Scenario \Rmnum{1}.

Now we consider the situation that both the measurements of BaBar and the contribution of $D_{sJ}(2700)$
measured by Belle are confirmed in the future.
We demonstrate that it is possible to reproduce about all measurements by using Scenario \Rmnum{2}:
a pole model with NR contribution in time-like form factors of $\bar D\bar K$, in addition.
Note that the first two categories share the same current-produced $\bar D\bar K$,
%which include NR terms,
while $\bar D^*\bar K$ form factors only appear in  Category 3 and 4.
Since modes in the last two categories already agree with data in Scenario \Rmnum{1}, using pole model only,
no NR contribution is introduced in $\bar D^*\bar K$ form factors.
The branching fractions of modes in the first two categories can be tuned by two complex NR parameters in
the time-like form factors of $\bar D\bar K$.
These two parameters are fixed by fitting
to the observed branching fractions of
$\Bbar_d \to D_d \Dbar^0K^-$ and $\Bbar_d \to D^*_d \Dbar^0K^-$ (denoted in the remarks in Table~\ref{table:threebody_exp_results}).
The best fit gives
$x^{DK}_{F_0}=(-75+52\text{i}){\rm GeV}^2$
and $x^{DK}_{F_1}=(16+2\text{i}){\rm GeV}^2$,
where $x^{DK}_{F_0}$ and $x^{DK}_{F_1}$ correspond to the NR contribution
in $\bar D\bar K$ time-like form factor $F_0$ and $F_1$, respectively [see Eq.~\eqref{eq:pole_and_non}].
%The best fitted NR contribution is rather large.
%This may due to the fact that the form of NR used is too naive or
%it reflects the tension in BaBar and Belle results in the fitting.
%A more sophisticated form of NR can be employed when more data is
%available and the discrepancy in BaBar and Belle measurements
%needed to be settled.
%
Usually the two complex (four real) NR parameters
cannot be fully determined from two constraints.
In this case, however, there is a localized and huge $D_{sJ}(2700)$ resonance
contribution in $\Bbar_d \to D^*_d \Dbar^0K^-$ mode.
The NR contribution, which is smooth in phase space, has
to cancel the $D_{sJ}(2700)$ contribution while maintaining the form factors
in other parts of phase space.
In other words, the phases of the NR parameters are constrained by the complex resonance,
while the magnitudes,
which control NR parts in the off-resonance region, are limited by data.
The branching fractions of the fit are shown
in Table~\ref{table:threebody_exp_results}, where 100\% uncertainties in $x\,$s are included in the first errors.
In this scenario,
all experimental results, except for the explicit disagreement
in $\Bbar_u \to D_u \Dbar^0K^-$ between data,
can be explained within uncertainty when NR is included.
In particular, the $\Bbar_d \to D^*_d \Dbar^0K^-$ rate is now reduced by a factor of 2 and consistent with data within errors.

%%%%%%%%%%%%%%%%%%%%%%%%%%%%%%%%%%%%%%%%%%%%%%
%% Table VI: Result of three-body states: Scenario 1
\begingroup
\squeezetable
\begin{table*}[t]
\begin{tabular}{b{1.4cm}  p{2.3cm} p{2.3cm} p{2.3cm} ||
    b{1.4cm}  p{2.3cm} p{2.3cm} p{2.3cm}}
\hline \hline
\multicolumn{8}{c}{Scenario \Rmnum{1}\,(\Rmnum{1}$^\prime$):}\\
\multicolumn{8}{c}{Pole Contribution Only}\\
\hline
\multicolumn{4}{c||}{Modes with $\Kbar$}
& \multicolumn{4}{c}{Modes with $\Kbar^*$} \\
\hline
Mode(f)  & $\mathcal{B}_{\Cur}(\Bbar_s \to f)(\%)$
         & $\mathcal{B}_{\Tr}(B_s \to f)(\%)$
         & $\DGamma_f / \Gamma_s(\%) $  &
Mode(f)  & $\mathcal{B}_{\Cur}(\Bbar_s \to f)(\%)$
         & $\mathcal{B}_{\Tr}(B_s \to f)(\%)$
         & $\DGamma_f / \Gamma_s (\%)$ \\
\hline
$D_s \Dbar^0K^-$
& \errr{0.19}{0.12}{0.04} \newline{(\errr{0.06}{0.03}{0.01})}
& \errr{0.04}{0.02}{0.01} \newline{(\errr{0.03}{0.02}{0.01})}
& \errr{0.17}{0.10}{0.03} \newline{(\errr{0.09}{0.04}{0.02})} &
$D_s \Dbar^0K^{*-}$
& (\errr{0.07}{0.03}{0.01})
& (\errr{0.03}{0.01}{0.01})
& (\errr{0.08}{0.04}{0.02})\\
$D_s D^-\Kbar^0$
& \errr{0.19}{0.12}{0.04} \newline{(\errr{0.05}{0.03}{0.01})}
& \errr{0.04}{0.02}{0.01} \newline{(\errr{0.03}{0.02}{0.01})}
& \errr{0.16}{0.09}{0.03} \newline{(\errr{0.08}{0.04}{0.02})} &
$D_s D^-\Kbar^{*0}$
& (\errr{0.06}{0.03}{0.01})
& (\errr{0.03}{0.01}{0.01})
& (\errr{0.08}{0.04}{0.02})\\
\hline
$D_s^* \Dbar^0K^-$
& \errr{0.64}{0.43}{0.13} \newline{(\errr{0.07}{0.03}{0.01})}
& \errr{0.09}{0.05}{0.02} \newline{(\errr{0.06}{0.03}{0.01})}
& \errr{0.38}{0.23}{0.08} \newline{(\errr{0.12}{0.05}{0.03})} &
$D_s^* \Dbar^0K^{*-}$
& (\errr{0.04}{0.02}{0.01})
& (\errr{0.03}{0.02}{0.01})
& (\errr{0.07}{0.03}{0.01})\\
$D_s^* D^-\Kbar^0$
& \errr{0.62}{0.42}{0.13} \newline{(\errr{0.07}{0.03}{0.01})}
& \errr{0.09}{0.05}{0.02} \newline{(\errr{0.06}{0.03}{0.01})}
& \errr{0.37}{0.22}{0.08} \newline{(\errr{0.11}{0.05}{0.02})} &
$D_s^* D^-\Kbar^{*0}$
& (\errr{0.04}{0.02}{0.01})
& (\errr{0.03}{0.02}{0.01})
& (\errr{0.07}{0.03}{0.02})\\
\hline
$D_s \Dbar^{*0}K^-$
& \errr{0.30}{0.18}{0.06} \newline{(\errr{0.17}{0.08}{0.04})}
& \errr{0.09}{0.05}{0.02} \newline{(\errr{0.08}{0.04}{0.02})}
& \errr{0.31}{0.21}{0.06} \newline{(\errr{0.23}{0.11}{0.05})} &
$D_s \Dbar^{*0}K^{*-}$
& (\errr{0.18}{0.08}{0.04})
& (\errr{0.08}{0.04}{0.02})
& (\errr{0.24}{0.12}{0.05})\\
$D_s D^{*-}\Kbar^0$
& \errr{0.29}{0.18}{0.06} \newline{(\errr{0.17}{0.08}{0.04})}
& \errr{0.09}{0.04}{0.02} \newline{(\errr{0.08}{0.04}{0.02})}
& \errr{0.30}{0.20}{0.06} \newline{(\errr{0.22}{0.11}{0.05})} &
$D_s D^{*-}\Kbar^{*0}$
& (\errr{0.17}{0.08}{0.04})
& (\errr{0.08}{0.04}{0.02})
& (\errr{0.24}{0.11}{0.05})\\
\hline
$D_s^* \Dbar^{*0}K^-$
& \errr{0.89}{0.59}{0.18} \newline{(\errr{0.14}{0.07}{0.03})}
& \errr{0.17}{0.09}{0.03} \newline{(\errr{0.11}{0.05}{0.02})}
& \errr{0.65}{0.39}{0.14} \newline{(\errr{0.23}{0.11}{0.05})} &
$D_s^* \Dbar^{*0}K^{*-}$
& (\errr{0.05}{0.02}{0.01})
& (\errr{0.04}{0.02}{0.01})
& (\errr{0.08}{0.04}{0.02})\\
$D_s^* D^{*-}\Kbar^0$
& \errr{0.86}{0.57}{0.18} \newline{(\errr{0.14}{0.06}{0.03})}
& \errr{0.16}{0.09}{0.03} \newline{(\errr{0.10}{0.05}{0.02})}
& \errr{0.64}{0.38}{0.13} \newline{(\errr{0.22}{0.10}{0.05})} &
$D_s^* D^{*-}\Kbar^{*0}$
& (\errr{0.05}{0.02}{0.01})
& (\errr{0.03}{0.02}{0.01})
& (\errr{0.08}{0.04}{0.02})\\
\hline
Total      &   &   & \errr{5.9}{3.6}{1.2}\footnotemark[1]
\newline{(\errr{2.6}{1.2}{0.5})\footnotemark[1]} &
Total      &   &   & (\errr{1.9}{0.9}{0.4})\footnotemark[1] \\
\hline \hline
\end{tabular}
\caption{The branching fractions ($\mathcal{B}_{\Cur,\Tr}$)
and width difference ($\DGamma_f$) of the
three-body $D^{(*)}_s\Dbar^{(*)}\Kbar^{(*)}$ modes in the scenario with
only pole contribution.
$\mathcal{B}_{\Cur}$ and $\mathcal{B}_{\Tr}$ denotes
the current-produced decay ($\Bbar_s \to f$) and
the transitional decay ($B_s \to f$), respectively.
$D_{sJ}(2700)$ is not included in modes with $\Kbar^*$ in this scenario.
The results with only $D^{(*)}_s$ poles are shown in parenthesis.
}
\footnotetext[1]{The contribution from $CP$ conjugate modes is included.}
\label{table:threebody_results1}
\end{table*}
\endgroup
%%%%%%%%%%%%%%%%%%%%%%%%%%%%%%%%%%%%%%%%%%%%%%%%%%%

\subsubsection{Branching Fractions in $B_{s}$ system and the Width Difference}

After checking the validity of our calculation by comparing to existing data on rates,
we move to our main purpose: estimating $\DGamma_s$.
The relevant diagram is shown in Fig.~\ref{fig:pole_diagram}.
In Table~\ref{table:threebody_results1}, we show our results
in Scenarios~\Rmnum{1}$^{(\prime)}$.
Recall that bounds on $\DGamma_s$ are related to rates [see Eq.~\eqref{eq:DGamma_limit_3}].
The branching fractions of current-produced modes and transition modes
are also shown, and can be read from $\mathcal{B}_{\Cur}(\Bbar_s \to f)$
and $\mathcal{B}_{\Tr}(B_s \to f)$, respectively.
For simplicity, only modes with $\Kbar^{(*)}$ are shown and
the results of modes with $K^{(*)}$ can be derived
from their $\CP$ conjugates.
As noted before, since $\CP$ is conserved in this work,
$\mathcal{B}(\Bbar_s \to f) = \mathcal{B}(B_s \to \bar{f})$ and
$\DGamma_f = \DGamma_{\bar{f}}$.
The total $\DGamma_f/\Gamma_s$ contains modes in the table
and their $\CP$ conjugates, so it is twice the sum of
the listed $\DGamma_f/\Gamma_s$ in the table.

Before discussing $\DGamma_s$,
we first look at branching fractions of these modes.
Current produced modes in $\Bbar_s$ decays are SU(3) related to modes considered previously.
Their rates are similar. % to the corresponding SU(3) related modes.
For example, $\bar B_s\to D_s^* \bar D^* K$ modes have largest
rates ($\sim0.88\%$) as the $\bar B_{u,d} \to D^*_{u,d} \bar D^* K$ modes.
However, the transition modes are new. Their rates are sub-percent or smaller.
Note that while current-produced modes with $\Kbar$ are dominated by $D_{sJ}(2700)$,
transition modes do not change significantly when $\bar D_{sJ}(2700)$ is included.
For instance, without $\bar D_{sJ}$ the branching fraction of current-produced mode
$\Bbar_s \to D_s^* \Dbar^0K^-$ drops from $0.64\%$ to  $0.07\%$.
In contrast, it drops only from $0.09\%$ to $0.06\%$ for
the branching fraction of transition mode $B_s \to D_s^* \Dbar^0K^-$.
The distinct behavior is not surprising because $B_s\to D^*_s\bar D_{sJ}(2700)$ rate (before $\bar D_{sJ}\to \Dbar^0K^-$) is relatively suppressed
compared with the $B_s \to D^*_s\bar D_s^*$ ones (before $\bar D_s^*\to \Dbar^0K^-$) (see Sec.~III.~A).
As we will see later, the different roles played by these poles
will be useful to enhance $\Delta\Gamma_s$ through interferences.

%%%%%%%%%%%%%%%%%%%%%%%%%%%%%%%%%%%%%%%%%%%%%%%%%%%%%%%%%
%% TableVII: Result of three-body states: Scenario 2
\begingroup
\squeezetable
\begin{table*}[t]
\begin{tabular}{b{1.4cm}  p{2.3cm} p{2.3cm} p{2.3cm} ||
    b{1.4cm}  p{2.3cm} p{2.3cm} p{2.3cm}}
\hline \hline
\multicolumn{8}{c}{Scenario \Rmnum{2}:}\\
\multicolumn{8}{c}{Pole contribution
  + NR in $\Dbar\Kbar$ time-like form factors}\\
\hline
\multicolumn{4}{c||}{Modes with $\Kbar$}
& \multicolumn{4}{c}{Modes with $\Kbar^*$} \\
\hline
Mode(f)  & $\mathcal{B}_{\Cur}(\Bbar_s \to f)(\%)$
         & $\mathcal{B}_{\Tr}(B_s \to f)(\%)$
         & $\DGamma_f / \Gamma_s(\%) $  &
Mode(f)  & $\mathcal{B}_{\Cur}(\Bbar_s \to f)(\%)$
         & $\mathcal{B}_{\Tr}(B_s \to f)(\%)$
         & $\DGamma_f / \Gamma_s (\%)$ \\
\hline
$D_s \Dbar^0K^-$
& ${0.09}^{+0.22}_{-0.02}\pm{0.02}$ & \errr{0.04}{0.02}{0.01} & \errr{0.08}{0.15}{0.01} &
$D_s \Dbar^0K^{*-}$
& (\errr{0.07}{0.03}{0.01})
& (\errr{0.03}{0.01}{0.01})
& (\errr{0.08}{0.04}{0.02})\\
$D_s D^-\Kbar^0$
& ${0.09}^{+0.22}_{-0.02}\pm{0.02}$ & \errr{0.04}{0.02}{0.01} & \errr{0.07}{0.13}{0.01} &
$D_s D^-\Kbar^{*0}$
& (\errr{0.06}{0.03}{0.01})
& (\errr{0.03}{0.01}{0.01})
& (\errr{0.08}{0.04}{0.02})\\
\hline
$D_s^* \Dbar^0K^-$
& ${0.31}^{+0.74}_{-0.13}\pm{0.13}$ & \errr{0.09}{0.05}{0.02} & \errr{0.11}{0.38}{0.02} &
$D_s^* \Dbar^0K^{*-}$
& (\errr{0.04}{0.02}{0.01})
& (\errr{0.03}{0.02}{0.01})
& (\errr{0.07}{0.03}{0.01})\\
$D_s^* D^-\Kbar^0$
& ${0.29}^{+0.71}_{-0.13}\pm{0.13}$ & \errr{0.09}{0.05}{0.02} & \errr{0.11}{0.38}{0.02} &
$D_s^* D^-\Kbar^{*0}$
& (\errr{0.04}{0.02}{0.01})
& (\errr{0.03}{0.02}{0.01})
& (\errr{0.07}{0.03}{0.02})\\
\hline
$D_s \Dbar^{*0}K^-$
& \errr{0.30}{0.18}{0.06} & \errr{0.09}{0.05}{0.02} & \errr{0.31}{0.21}{0.06} &
$D_s \Dbar^{*0}K^{*-}$
& (\errr{0.18}{0.08}{0.04})
& (\errr{0.08}{0.04}{0.02})
& (\errr{0.24}{0.12}{0.05})\\
$D_s D^{*-}\Kbar^0$
& \errr{0.29}{0.18}{0.06} & \errr{0.09}{0.04}{0.02} & \errr{0.30}{0.20}{0.06} &
$D_s D^{*-}\Kbar^{*0}$
& (\errr{0.17}{0.08}{0.04})
& (\errr{0.08}{0.04}{0.02})
& (\errr{0.24}{0.11}{0.05})\\
\hline
$D_s^* \Dbar^{*0}K^-$
& \errr{0.89}{0.59}{0.18} & \errr{0.17}{0.09}{0.03} & \errr{0.65}{0.39}{0.14} &
$D_s^* \Dbar^{*0}K^{*-}$
& (\errr{0.05}{0.02}{0.01})
& (\errr{0.04}{0.02}{0.01})
& (\errr{0.08}{0.04}{0.02})\\
$D_s^* D^{*-}\Kbar^0$
& \errr{0.86}{0.57}{0.18} & \errr{0.16}{0.09}{0.03} & \errr{0.64}{0.38}{0.13} &
$D_s^* D^{*-}\Kbar^{*0}$
& (\errr{0.05}{0.02}{0.01})
& (\errr{0.03}{0.02}{0.01})
& (\errr{0.08}{0.04}{0.02})\\
\hline
Total      &   &   & \errr{4.5}{4.4}{0.9}\footnotemark[1] &
Total      &   &   & (\errr{1.9}{0.9}{0.4})\footnotemark[1] \\
\hline \hline
\end{tabular}
\caption{The branching fractions ($\mathcal{B}_{\Cur,\Tr}$) and
width difference ($\DGamma_f$) of the
three-body $D^{(*)}_s\Dbar^{(*)}\Kbar^{(*)}$ modes in Scenario \Rmnum{2} where
$\Dbar \Kbar$ time-like form factors have NR contribution.
The notation is the same as in Table~\ref{table:threebody_results1}.
}
\footnotetext[1]{The contribution from $CP$ conjugate modes is included.}

\label{table:threebody_results2}
\end{table*}
\endgroup
%%%%%%%%%%%%%%%%%%%%%%%%%%%%%%%%%%%%%%%%%%%%%%%%%%%%

As the branching fractions of transition modes are not tiny,
one would expect a non-negligible $\DGamma_s$.
The $\DGamma_f / \Gamma_s$ of three-body modes range from $0.07\%$ to $0.65\%$ as shown in Table~VI.
The last two modes with $\Kbar$ have the largest $\DGamma_f$ as their rates are largest.
In this scenario,
the total $\DGammas / \Gamma_s$ is
%% Total DGamma Sce. 1
\be
\label{eq:threebodyresult_1}
\DGammas/ \Gamma_s
{(D^{(*,**)}_s \Dbar^{(*,**)}_s)}  &= & (10.4 \pm 2.5 \pm 2.2)\%,
\non\\
\DGammas/ \Gamma_s {(D^{(*)}_s \bar{D}^{(*)}\Kbar+\bar D^{(*)}_s {D}^{(*)}K)} & = & (5.9 \pm 3.6 \pm 1.2)\%,
\non\\
\DGammas/ \Gamma_s {(D^{(*)}_s \bar{D}^{(*)}\Kbar^*+\bar D^{(*)}_s {D}^{(*)}K^*)}  &= & (1.9 \pm 0.9 \pm 0.4)\%,  \\
\DGammas/ \Gamma_s   &= & (18.2 \pm 7.0 \pm 3.8) \%.
\non
\en
Clearly, the $\DGammas$ of three-body modes is comparable to two-body modes.
The $\DGammas$ of three-body modes is mainly comprised of modes with $K$.
It shows that the approximation in which $D^{(*)}_s\Dbar^{(*)}_s$ modes
saturate $\DGamma_s$ is dubious.
In addition, Eq.~\eqref{eq:threebodyresult_1} agrees with
the short-distance calculation in Eq.~(\ref{eq:DGamma_sd})
within uncertainties.
There is no evidence of the violation of short-distance result
and the underlying OPE assumption.

The interference between $\bar D_{sJ}(2700)$ and $\bar D^{(*)}_s$ can be studied
by comparing Scenario \Rmnum{1} with Scenario \Rmnum{1}$^\prime$ and the result of $\bar D_{sJ}(2700)$.
The full treatment of modes with $\Kbar$ in Scenario \Rmnum{1},
where $\bar D_{sJ}(2700)$ and $\bar D^{(*)}_s$ are taken into consideration simultaneously,
gives $\DGamma_s/\Gamma_s\simeq 5.9\%$.
On the other hand,
one can treat $\bar D_{sJ}(2700)$ and $\bar D^{(*)}_s$ separately and
sum their $\DGamma_s/\Gamma_s$.
The contribution of $\bar D^{(*)}_s$ only (Scenario \Rmnum{1}$^\prime$)
can be read from the Table.
For $\bar D_{sJ}(2700)$, its contribution can be estimated from
the two-body calculation (see Sec.~III~A) with narrow width approximation.
We further check that it decreases from the two-body result of $1.9\%$ to $1.7\%$,
when full three-body calculation is imposed.
In the case that $\bar D_{sJ}(2700)$ and $\bar D^{(*)}_s$ are sum separately,
the total $\DGamma_s/\Gamma_s$ of modes with $\Kbar$
is only $2.6\%+1.7\%=4.3\%$, smaller than $5.9\%$ in Scenario \Rmnum{1}.
The difference, which is about the size of the $\bar D_{sJ}$ contribution alone,
shows that there is considerable interference
between $\bar D_{sJ}(2700)$ and $\bar D^{(*)}_s$ poles.
Such interference can be understood as followes.
As depicted in Fig.~\ref{fig:pole_diagram},
the $\Dbar^{(*)}\Kbar^{(*)}$ pairs emitted by the current-produced $\Dbar_{sJ}(2700)$ pole
interfere with the same states from the transited $\bar D^{(*)}_s$ poles in transition diagram.
Unlike the highly suppressed $B_s\to \bar D_{sJ}$ transitions (see Table~\ref{table:DsJ_twobody}),
the $B_s\to \bar D^{(*)}_s$ transitions are sizable (see Table~\ref{table:twobody_results}),
leading to enhanced $\Bbar_s-B_s$ mixing and $\DGamma_s$.
In short, $\DGamma_s$ receives the interference contribution from
current-produced $\Dbar_{sJ}(2700)$ pole (from $\bar B_s$ decays)
and transited $\bar D_s^{(*)}$ poles (from $B_s$ decays),
which bypass the mismatch of current-produced and transited $\Dbar_{sJ}$ in two-body modes.
In total, diagrams containing $\bar D_{sJ}$ pole contribute more than those with $\bar D^{(*)}_s$ poles only.

One can bound the width difference in Table~\ref{table:threebody_results1}
by Eq.~\eqref{eq:DGamma_limit_3}.
For example, the $\DGamma_f/\Gamma_s$ is bounded to be $0.77\%$ and
$0.08\%$ for $D_s^* \Dbar^{*0}K^-$ and $D_s^* \Dbar^{*0}K^{*-}$ modes, respectively.
Comparing to $\DGamma_f$, we see that the bounds in modes with $\Kbar$ are higher within $20\%$,
while they constrain $\DGamma_f$  very well for modes with $\Kbar^*$.
The accuracy of $\DGamma_f$ estimation in modes with $\Kbar^*$
has to do with the virtual $\bar D^{(*)}_s$ poles.
The pole contribution of $\bar D^{(*)}_s$ is almost real and so are the
resulting amplitudes.
As a result, the suppression from
the inequality of Eq.~(\ref{eq:DGamma_limit_1})
is tiny for modes with $\Kbar^*$.
This demonstrates that the virtual $\bar D^{(*)}_s$ poles are very
efficient to mediate the width difference.
On the contrary, the on-shell $\bar D_{sJ}(2700)$, which plays an important role
in modes with $\Kbar$, generates complex amplitudes and result in the
suppression of $\DGamma_f$ in these modes.

The results of Scenario \Rmnum{2}
are shown in Table~\ref{table:threebody_results2}.
Only the first four modes with $\Kbar$ are different
from Scenario \Rmnum{1}.
Note that all transition modes and modes with $\Kbar^*$ are still the same as
in Scenario \Rmnum{1}, since there is no measurement at all to call beyond pole model.
One can read from the table that the $\DGamma_f$ of the first four modes (modes with NR)
drop by 50\% to 70\%.
The decrease is caused by the reduction of the branching fractions in
current-produced modes.
Morever, the actual $\DGamma_f$ moves away from the upper bound
in Eq.~\eqref{eq:DGamma_limit_3} when the complex NR contribution are included.
In this scenario, the total $\DGammas / \Gamma_s$ is
%% Total \DGamma Sce. 2
\be
 \label{eq:threebodyresult_2}
\DGammas/ \Gamma_s
{(D^{(*,**)}_s \Dbar^{(*,**)}_s)} & = & (10.4 \pm 2.5 \pm 2.2)\%,
\non\\
\DGammas/ \Gamma_s {(D^{(*)}_s \bar{D}^{(*)}\Kbar+\bar D^{(*)}_s {D}^{(*)}K)}  &= & (4.5 \pm 4.4 \pm 0.9)\%,
\non\\
\DGammas/ \Gamma_s {(D^{(*)}_s \bar{D}^{(*)}\Kbar^*+\bar D^{(*)}_s {D}^{(*)}K^*)}  &= & (1.9 \pm 0.9 \pm 0.4)\%,
\\
\DGammas/ \Gamma_s   &= & (16.7 \pm 7.8 \pm 3.5) \%.
\non
\en
Despite the drop of $\DGamma_f$ in modes with NR,
the total $\DGammas$ remains similar to Scenario \Rmnum{1}
because these modes are not dominant in $\DGammas$.
Most features are similar to the previous case.
The effect of three-body modes is still non-negligible.
It is interesting to see that the central value is
more consistent to short-distance calculation.
The conclusion remains the same as in Scenario \Rmnum{1}.

%%%%%%%%%%%%%%%%%%%%%%%%%%%%%%%%%%%%%%%%%%%%%%%%%%%%%%%%%
\section{Discussion}\label{sec:discussion}
%%%%%%%%%%%%%%%%%%%%%%%%%%%%%%%%%%%%%%%%%%%%%%%%%%%%%%%%%

%%%%%%%%%%%%%%%%%%%%%%%%%%%%%%%%%%%%%%%%%%%%%%%
%% TableVIII: Result of three-body states: Scenario 3
\begingroup
\squeezetable
\begin{table*}[t]
\begin{tabular}{b{1.4cm}  p{2.3cm} p{2.3cm} p{2.3cm} ||
    b{1.4cm}  p{2.3cm} p{2.3cm} p{2.3cm}}
\hline \hline
\multicolumn{8}{c}{Scenario \Rmnum{3}:} \\
\multicolumn{8}{c}{$D_{sJ}(2700)$ is included in all modes}\\
\hline
\multicolumn{4}{c||}{Modes with $\Kbar$}
& \multicolumn{4}{c}{Modes with $\Kbar^*$} \\
\hline
Mode(f)  & $\mathcal{B}_{\Cur}(\Bbar_s \to f)(\%)$ \newline(exp.)
         & $\mathcal{B}_{\Tr}(B_s \to f)(\%)$
         & $\DGamma_f / \Gamma_s(\%) $ \newline(limit) &
Mode(f)  & $\mathcal{B}_{\Cur}(\Bbar_s \to f)(\%)$
         & $\mathcal{B}_{\Tr}(B_s \to f)(\%)$
         & $\DGamma_f / \Gamma_s (\%)$ \newline(limit) \\
\hline
$D_s \Dbar^0K^-$
& ${0.09}^{+0.22}_{-0.02}\pm{0.02}$ & \errr{0.04}{0.02}{0.01} & \errr{0.08}{0.15}{0.01} &
$D_s \Dbar^0K^{*-}$
& \errr{0.10}{0.05}{0.02} & \errr{0.03}{0.02}{0.01} & \errr{0.11}{0.05}{0.02}\\
$D_s D^-\Kbar^0$
& ${0.09}^{+0.22}_{-0.02}\pm{0.02}$ & \errr{0.04}{0.02}{0.01} & \errr{0.07}{0.13}{0.01} &
$D_s D^-\Kbar^{*0}$
& \errr{0.09}{0.05}{0.02} & \errr{0.03}{0.01}{0.01} & \errr{0.10}{0.05}{0.02}\\
\hline
$D_s^* \Dbar^0K^-$
& ${0.31}^{+0.74}_{-0.13}\pm{0.13}$ & \errr{0.09}{0.05}{0.02} & \errr{0.11}{0.38}{0.02} &
$D_s^* \Dbar^0K^{*-}$
& \errr{0.27}{0.13}{0.06} & \errr{0.06}{0.03}{0.01} & \errr{0.23}{0.11}{0.05}\\
$D_s^* D^-\Kbar^0$
& ${0.29}^{+0.71}_{-0.13}\pm{0.13}$ & \errr{0.09}{0.05}{0.02} & \errr{0.11}{0.38}{0.02} &
$D_s^* D^-\Kbar^{*0}$
& \errr{0.25}{0.12}{0.06} & \errr{0.05}{0.02}{0.01} & \errr{0.21}{0.10}{0.04}\\
\hline
$D_s \Dbar^{*0}K^-$
& \errr{0.30}{0.18}{0.06} & \errr{0.09}{0.05}{0.02} & \errr{0.31}{0.21}{0.06} &
$D_s \Dbar^{*0}K^{*-}$
& \errr{0.28}{0.13}{0.06} & \errr{0.10}{0.05}{0.02} & \errr{0.32}{0.16}{0.07}\\
$D_s D^{*-}\Kbar^0$
& \errr{0.29}{0.18}{0.06} & \errr{0.09}{0.04}{0.02} & \errr{0.30}{0.20}{0.06} &
$D_s D^{*-}\Kbar^{*0}$
& \errr{0.27}{0.13}{0.06} & \errr{0.09}{0.04}{0.02} & \errr{0.31}{0.15}{0.07}\\
\hline
$D_s^* \Dbar^{*0}K^-$
& \errr{0.89}{0.59}{0.18} & \errr{0.17}{0.09}{0.03} & \errr{0.65}{0.39}{0.14} &
$D_s^* \Dbar^{*0}K^{*-}$
& \errr{0.23}{0.11}{0.05} & \errr{0.05}{0.03}{0.01} & \errr{0.21}{0.10}{0.04}\\
$D_s^* D^{*-}\Kbar^0$
& \errr{0.86}{0.57}{0.18} & \errr{0.16}{0.09}{0.03} & \errr{0.64}{0.38}{0.13} &
$D_s^* D^{*-}\Kbar^{*0}$
& \errr{0.21}{0.10}{0.04} & \errr{0.05}{0.03}{0.01} & \errr{0.20}{0.09}{0.04}\\
\hline
Total      &   &   & \errr{4.5}{3.0}{0.9}\footnotemark[1] &
Total      &   &   & \errr{3.4}{1.6}{0.7}\footnotemark[1] \\
\hline \hline
\end{tabular}
\caption{The branching fractions ($\mathcal{B}_{\Cur, \Tr}$) and
width difference ($\DGamma_f$) of the
three-body $D^{(*)}_s\Dbar^{(*)}\Kbar^{(*)}$ modes in Scenario \Rmnum{3},
where $D_{sJ}(2700)$ is included in all modes.
The notation is the same as in Table~\ref{table:threebody_results1}.}
\footnotetext[1]{The contribution from $CP$ conjugate modes is included.}
\label{table:threebody_results3}
\end{table*}
\endgroup
%%%%%%%%%%%%%%%%%%%%%%%%%%%%%%%%%%%%%%%%%%%%

We have seen that $\bar D_{sJ}(2700)$ is important in modes with $\bar K$.
One expects $\bar D_{sJ}(2700)$ to be non-negligible in modes with $\bar \Kst$ as well.
Even though $\bar D_{sJ}(2700)$ is not heavy enough to decay to on-shell
$\bar D^{(*)}\bar\Kst$, its width is wide and its mass is close
to the invariant mass threshold of $\bar D^{(*)}\bar\Kst$.
Unfortunately, there is no information about the coupling constants of the
effective Lagrangian for $\bar D_{sJ}(2700) \to \bar D^{(*)}\bar\Kst$.
Unlike the on-shell $\bar D_{sJ}(2700) \to \bar D^{(*)}\bar K$ decay,
we cannot extract
the coupling constant of $\bar D_{sJ}(2700) \to \bar D^{(*)}\bar \Kst$ directly from data.
Thus, for illustration, we set the coupling constants in analogy
to the coupling constants of $\bar D^*$ to $\bar D^{(*)}\bar K^{(*)}$ vertices
%% Coupling constant in analogy with DK/DK*
\begin{equation}
 \tilde{g}_{D_{sJ}D^{(*)}\Kst}
   \approx \tilde{g}_{D_{sJ}D^{(*)}K} (\frac{g_{D^*D^{(*)}\Kst}}{g_{D^*D^{(*)}K}})
   \approx 0.5 \times \tilde{g}_{D_{sJ}D^{(*)}K}.
 \label{eq:DsJKst_coup}
\end{equation}
Table~\ref{table:threebody_results3} shows the result in this analogy,
which we call Scenario~\Rmnum{3}.
The results of modes with $\Kbar$ remain the same as in Scenario~\Rmnum{2}.

Comparing with the results in previous scenarios,
all branching fractions and $\DGamma_f$ increase.
As before, the effect of $\bar D_{sJ}(2700)$ is stronger in
current-produced modes than in transition modes.
In particular, current-produced modes in Category 2 ($D^*_s\bar D\bar K^*$)
and 4 ($D^*_s\bar D^*\bar K^*$) are very sensitive
to the appearance of $\bar D_{sJ}(2700)$.
Their branching fractions rise at least four times.
This large effect of current-produced $D_{sJ}(2700)$ in Category 2 and 4
is similar to modes with $\Kbar$.
If there is a measurement of modes in these two categories,
it is possible to extract $\tilde{g}_{D_{sJ}D^{(*)}\Kst}$ by fitting to branching
fractions.
The $\tilde{g}_{D_{sJ}D^{(*)}\Kst}$ in return could help the identification of
$D_{sJ}(2700)$.
The current-produced modes with $\Kbar^*$ have branching fractions in the
order of $10^{-3}$, similar to modes with $\Kbar$.

The rise of branching fractions in current-produced modes lead to
the increase of $\DGammas$.
Following the trend of branching fractions,
$\DGamma_f$ in Category 2 and 4 have significant increase compared with the other two.
In this scenario, the total $\DGammas / \Gamma_s$ is
%% Total \DGamma Sce. 3
\be
\label{eq:threebodyresult_3}
\DGammas/ \Gamma_s
{(D^{(*,**)}_s \Dbar^{(*,**)}_s)}  &= & (10.4 \pm 2.5 \pm 2.2)\%,
\non\\
\DGammas/ \Gamma_s {(D^{(*)}_s \bar{D}^{(*)}\Kbar+\bar D^{(*)}_s {D}^{(*)}K)}  &= & (4.5 \pm 4.4 \pm 0.9)\%,
\non\\
\DGammas/ \Gamma_s {(D^{(*)}_s \bar{D}^{(*)}\Kbar^*+\bar D^{(*)}_s {D}^{(*)}K^*)}  &= & (3.4 \pm 1.6 \pm 0.7)\%,
\\
\DGammas/ \Gamma_s  & = & (18.2 \pm 8.5 \pm 3.8) \%.
\non
\en
The total $\DGamma_s$ induced by modes with $\Kbar^*$ almost doubles.
The effect from three-body modes is strengthen by considering the off-shell
decay of $\bar D_{sJ}(2700)$ to $\bar D^{(*)}\bar \Kst$.
For total $\DGamma_s$, the central value returns to the one
in Scenario \Rmnum{1}.
Total $\DGamma_s$ does not alter significantly as the contribution for
modes with $\Kbar^*$ is not dominant.
The result still agrees with short-distance calculation.

The interference in modes with $\Kbar^*$ is strong.
Similar to the discussion in Scenario \Rmnum{1},
if we leave only $\bar D_{sJ}(2700)$ and turn off $\bar D^{(*)}_{s}$ poles,
the resulting $\DGamma_f / \Gamma_s$ of these modes is only $0.3\%$.
It is much smaller than the $1.5\%$ increase found in Scenario \Rmnum{3} (compared to Scenario \Rmnum{2}).
Recalling the result in Scenario \Rmnum{1},
one finds that modes with $\Kbar^*$ allow more constructive interference
than modes with $\Kbar$.
For modes with $\Kbar$, the interference is restricted by the on-shell
$\bar D_{sJ}(2700)$ resonance, which is localized in phase space.
On the contrary,
the $\bar D_{sJ}(2700)$ resonance becomes off-shell and hence smooth in phase space
for modes with $\Kbar^*$.
It is more coherent to the $\bar D^{(*)}_{s}$ pole contributions
and interfere with them better.
As in the $\Kbar$ case,
the interference, mediated by the $\bar D^{(*)}\bar K^{*}$ pair, is comparable to
the contribution of $\bar D_{sJ}(2700)$ itself.

We show that the branching fractions of these modes are
in the order of $10^{-3}$ to $10^{-4}$.
Recall that there is no corresponding measurement
in current-produced modes with $\bar \Kst$ and in all transition modes.
For current-produced modes with $\bar \Kst$,
they can be studied in $\bar B_{u,d}$ system in analogy to modes with $\bar K$.
These branching fractions should be measurable with current
data collected by the B factories.
On the other hand, $\bar B_{u,d}$ systems have more different behaviors
in transition modes.
$B_{u,d}$ transit to $\Dbar^{(*)}{\pi}$ pairs
instead of $\Dbar^{(*)}\Kbar^{(*)}$.
The $\Dbar^{(*)}{\pi}$ pairs can be produced either from
nearly on-shell $\Dbar^{*}$
or from other on-shell intermediate resonances.
One expects the transition modes in $B_{u,d}$ are enhanced than in $B_s$.
In fact, semileptonic modes with $B_{u,d} \to \Dbar^{(*)}{\pi}_{u,d}$
transition have been measured~\cite{PDG}.
The branching fractions are around $0.5\%$,
much larger than the transition modes in this work.
For the purpose of estimating the width difference,
$\DGammas$ can be bounded by Eq.~\eqref{eq:DGamma_limit_3}
when current-produced and transition modes are measured.
Independent of $\DGammas$, experimental studies of these modes
will be interesting enough in their own right.

So far we fit the decay constant of $D_{sJ}(2700)$ by its contribution
to $\Bbar_u \to D_u \Dbar^0K^-$ as measured by Belle.
If future experiments favor the result of BaBar and lower the contribution
of $D_{sJ}(2700)$, then the decay constant will be smaller.
In such case, the branching fractions of modes in Category 1 and 2
in pole model become smaller and may be consistent with experiments without resorting to NR contribution.
Nevertheless, the branching fractions of modes in Category 3 and 4
will be deficient. Similar to Scenario 2, one can then
add NR contribution in the time-like form factors of $\bar\Dst \bar K$
to fit the observed branching fractions.
Although there are more NR parameters in $\bar\Dst \bar K$ form factors,
one can extract information in the Dalitz plots, especially the interference
between the continuum and the $D_{sJ}(2700)$ resonance.
These can be studied after future measurements are done.

In principle, modes with $s\bar{s}$,
such as $\eta^{(\prime)}$, $\omega$, and $\varphi$,
can also contribute to $\DGamma_s$.
These modes are difficult to calculate because they mix current-produced,
transition, and color-suppressed diagrams together.
Nonetheless, we find that the contribution of these modes are small.
The phase space is suppressed and the number of modes are less.
We estimate the contribution to $\DGamma_s$ by
$D^{(*)}_s\Dbar^{(*)}_s\eta^{(\prime)}$ modes with color-allowed diagrams only.
The effect is less than $0.7\%$, which is negligible.

\begin{figure}[t]
\centering
\subfigure[4-body diagram type \Rmnum{1}]{
  \includegraphics[width=5cm]{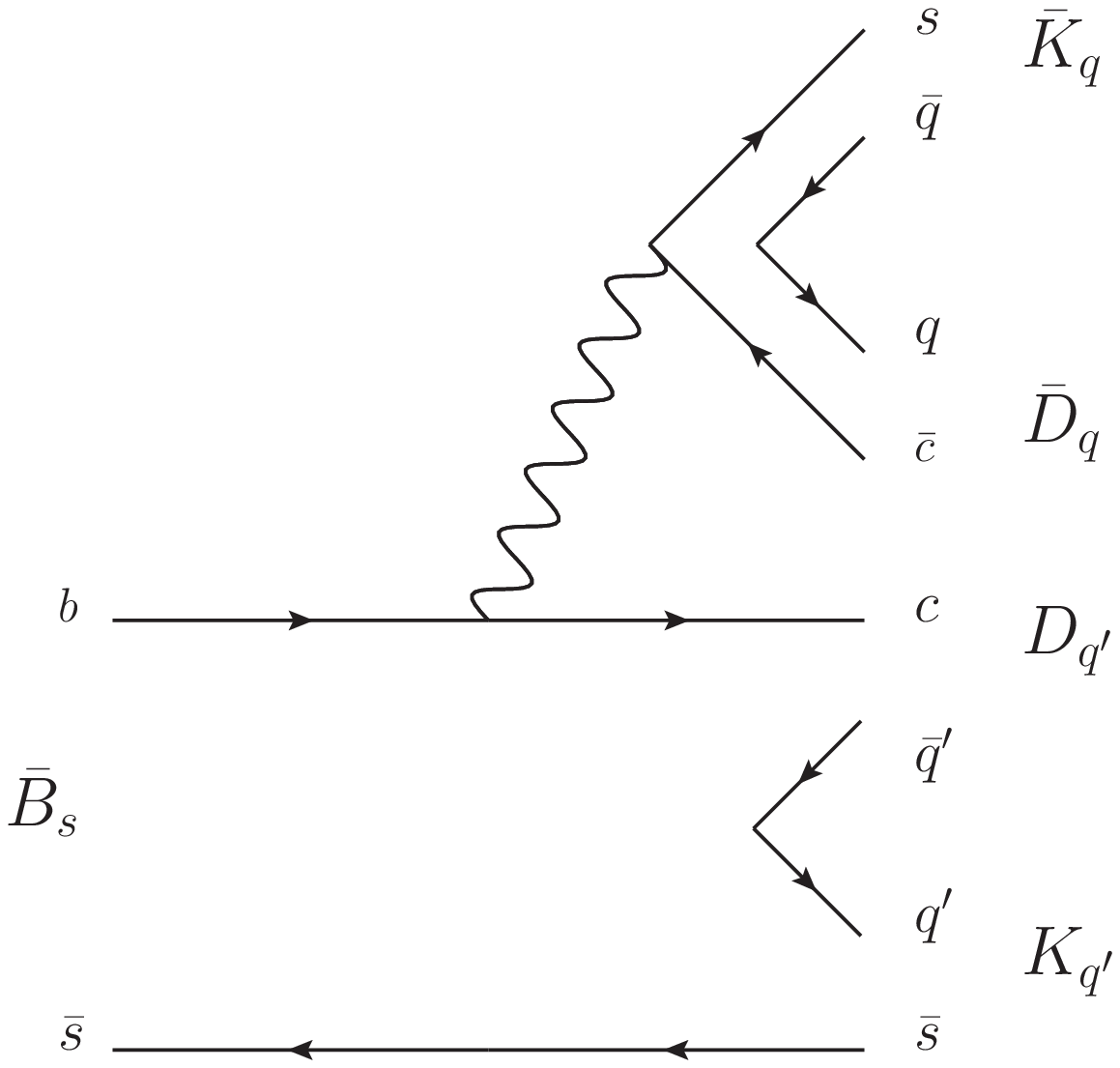}
}
\subfigure[4-body diagram type \Rmnum{2}]{
  \includegraphics[width=9.5cm]{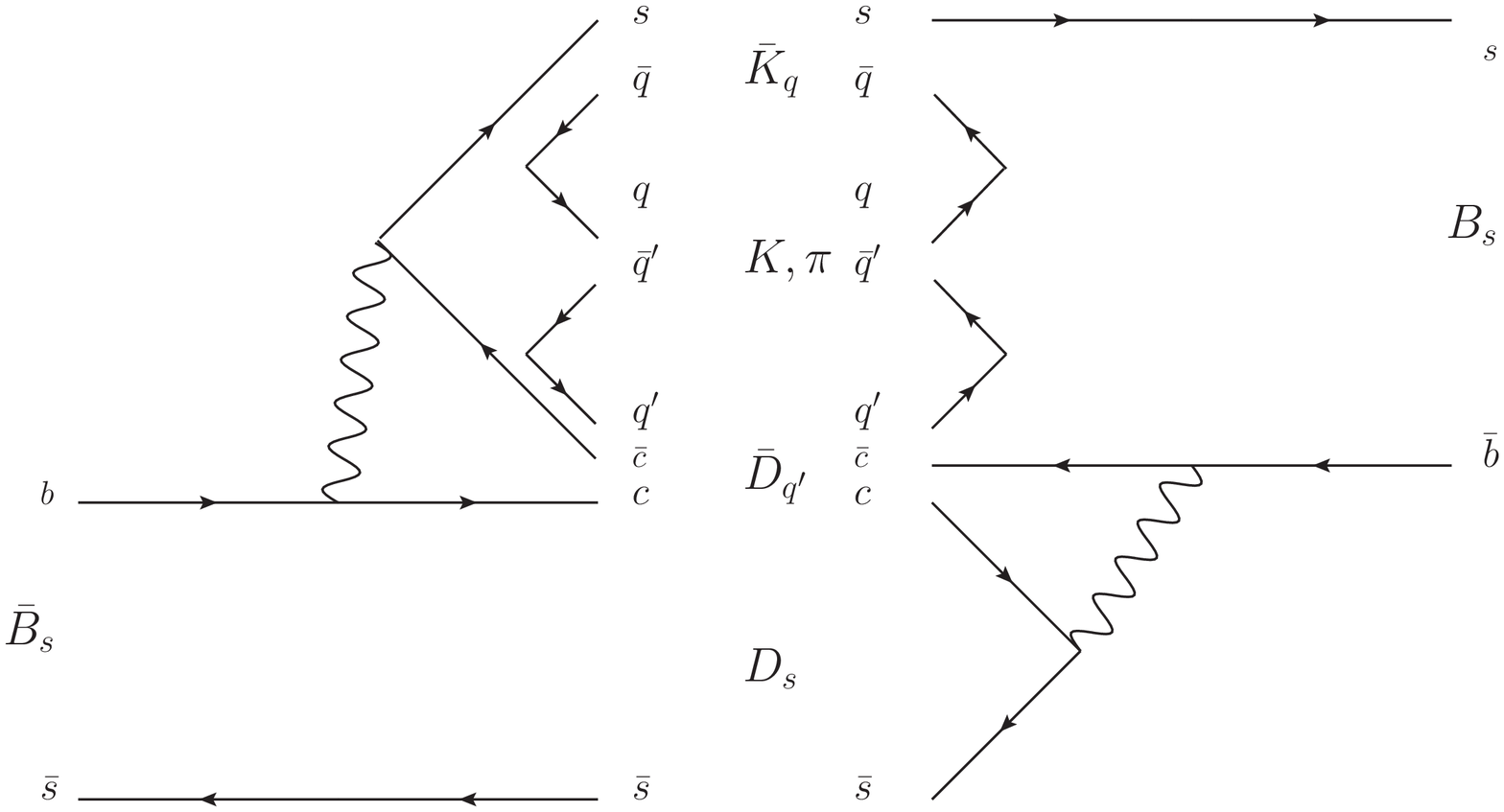}
}
\caption{Left: The first type 4-body diagram. Both of current and transition
  part have an $q\bar{q}$ pair.
  Right: The second type 4-body diagrams.
  All $q\bar{q}$ pairs lies in current,
  or in transition for $B_s$ decays to this mode.}
\label{fig:4body}
\end{figure}

We have shown that the effect of three-body modes could be sizable.
It is interesting to
see if other high-order modes could have similar effect on $\DGamma_s$.
Note that the phase space is gradually saturated from $D_s\Dbar \bar K$ mode to
$D^*_s\Dbar^*\bar K^*$ mode, and
the effect of high-order modes may be limited.
Fig.~\ref{fig:4body} shows the diagrams of possible four-body modes.
The first type of diagram (left diagram in in Fig.~\ref{fig:4body})
can produce $D^{(*)}\Dbar^{(*)} K^{(*)} \Kbar^{(*)}$,
but the two $K$ mesons
cannot be simultaneously in excited states because of insufficient phase space.
The amplitude of this diagram can be calculated
with the same form factors as in three-body modes.
We roughly estimate the branching fraction of this type of diagrams,
which is two orders of magnitude smaller than three-body modes.
Given that the number of $D^{(*)}\Dbar^{(*)} K^{(*)} \Kbar^{(*)}$ modes
is 48, only $0.5$ times more than 3-body diagrams,
the contribution of these diagrams are still negligible.
The second type may involve pions and could have a larger phase space.
We calculate the dimensionless fraction of phase space area
\begin{equation}
\frac{1}{m_B^2}
  \frac{\mathcal{A}^{\Phi}(\text{4-body})}{\mathcal{A}^{\Phi}\text{(3-body)}} < 10^{-4},
\end{equation}
where $\mathcal{A}^{\Phi}$ is the phase space area.
This ratio strongly suggests that the effect of 4-body modes is negligible.
Even if the branching fractions of current amplitudes
are large, the branching fraction of
transition diagrams may not be as large as in current amplitudes.
It should be safe to estimate $\DGamma_s$ up to three-body modes.
%% 4-body diagrams

%%%%%%%%%%%%%%%%%%%%%%%%%%%%%%%%%%%%%%%%%%%%%%%%%%%%%%%%%
\section{Conclusion}\label{sec:conclusion}
%%%%%%%%%%%%%%%%%%%%%%%%%%%%%%%%%%%%%%%%%%%%%%%%%%%%%%%%%

In conclusion, we have estimated the long-distance contribution to
$\DGamma_s$ of the $B_s-\Bbar_s$ system.
First,
we revisit the contributions by two-body $D^{(*)}_s\Dbar^{(*)}_s$ modes.
The $\DGamma_s/\Gamma_s$ by these modes is $(10.2\pm2.2\pm2.1)\%$,
which decreases from previous result in Ref.~\cite{Aleksan:1993qp}.
More precise measurements in $B_s$ system can help
extract more accurate parameters and improve the theoretical prediction.
After including $D_{s0}^*(2317)$, $D_{s1}(2460)$, and $D_{s1}(2536)$ resonances,
the $\DGamma_s/\Gamma_s$ change only slightly to ($10.4\pm2.5\pm2.2)\%$.

For the three-body $D^{(*)}_s \Dbar^{(*)} \Kbar^{(*)}$ modes,
factorization formalism with form factors modeled by
$\bar D_s^{(*)}$ and $\bar D_{sJ}(2700)$ poles and non-resonant (NR) contributions, if necessary, are used.
The branching fractions predicted by pole models are consistent with experiment
in two of the four categories, while the agreement in the remaining modes with data can be achieved
by including NR contribution.
Three-body modes can bypass some difficulties in two body modes.
In particular, sizable constructive interference between $\bar D_{sJ}$ and $\bar D^{(*)}_s$ poles,
which is impossible for two body modes,  are found.

Our results on $\DGamma_s$ in three scenarios are summarized
in Eq.~\eqref{eq:threebodyresult_1}, \eqref{eq:threebodyresult_2},
and \eqref{eq:threebodyresult_3}.
Although the three scenarios have different theoretical assumptions,
it is of interest to note that the resulting $\DGammas$ values are similar.
Thus, we give the following concluding remarks.
First, the total $\DGamma_s$ agrees with short-distance calculation.
In other words, long distance contributions from $b \to c\bar{c}s$ decays
do not enhance $\DGamma_s$ (or the real part of $\Gamma_{12,s}$) significantly.
This demonstrates
that the short-distance result and the assumption of OPE are reliable.
If the anomalous dimuon asymmetry with sizable $\Delta \Gamma_s$ is confirmed in the future,
the enhancement in $\DGamma_s$ must have origins from new physics.

Second, we find that the effect of three-body modes ($\sim 8\%$) is
comparable to two-body modes ($\sim 10\%$).
The assumption that two-body decays saturate $\DGamma_s$,
%which is true under the limits ($(m_b-2m_c) \to 0$ and $m_c \to \infty$)
%and $N_c \to \infty$,
receives a considerable correction.
This correction comes from both $D_{sJ}(2700)$ and off-shell $D^{(*)}_s$ poles.

We end our conclusion by pointing out some experimental issues
where progress can be made in the near future.
Two body modes in $B_s$ decays need to be measured with better precisions (see Sec.~III.~A).
For three body modes, up to now, there is no measurement of transition modes,
nor on modes with $\Kst$ in $B_{u,d}$ system.
Even the available measurements in current-produced modes with $K$
contain inconsistencies.
In particular,  the $2.2\sigma$ difference between Belle and BaBar
in $B^- \to D^0 \Dbar^0K^-$ mode has to be resolved.
From Tables in Sec. III and IV, we see that many modes
remain to be found or confirmed experimentally.
For example, $\bar B_s\to D^*_s \bar D^{(*)} \bar K^{(*)}$ rates are predicted
at the percent level and can be observed soon.
Note that the modes with $\bar D^{(*)}\bar K^*$ will be useful to
extract the $D_{sJ}$ strong coupling.
Although the measurements of two and three-body decay rates are useful for
refining the theoretical prediction
and to set bound on $\Delta\Gamma_s$, these modes are of interest in their own right.
We hope that (Super-) B factories and LHCb can complete the measurements of
these missing modes.

\vskip0.25cm
\noindent \textbf{Note Added.} 
After the completion of this paper, we noticed the work of
Ref.~\cite{Lenz:2011zz}, which pointed out that penguin
contributions to $B_s \to J/\psi \phi$ could reduce somewhat 
the need for enhanced $\Delta \Gamma_s$. It also reiterates
the point made in the second reference of Ref.~\cite{Lenz:2007JHEP}
that there is no indication of large or ill-behaved corrections 
to the short distance expansion (or Heavy Quark Expansion).

\begin{acknowledgements}
CKC thanks the support by National Science Council of Taiwan under
grant number: NSC-97-2112-M-033-002-MY3 and NSC-100-2112-M-033-001-MY3.
WSH and CHS are grateful to the National Science Council of Taiwan
for the support of the Academic Summit grant, NSC 99-2745-M-002-002-ASP.
\end{acknowledgements}

%%%%%%%%%%%%%%%%%%%%%%%%%%%
%%%%%%%%%%%%%%%%%%%%%%%%%%%
\appendix

\section{Basic decay constants and form factors}
\label{basic_parameters}
The value of basic parameters are summarized in this section.
We take Wilson coefficients $c_1=1.081$ and $c_2=-0.190$
with naive factorization. This corresponds to
%% a1
\begin{equation}
  a_1 = 1.02 \pm 0.10,
  \label{eq:a1}
\end{equation}
where we estimate a $10\%$ uncertainty.
The decay constants of $D_{u,d}$ and form factors of $\Bbar_{u,d} \to D_{u,d}$
are given in Ref.~\cite{Cheng:2003sm}.

For calculating $\Bbar_s \to D^{(*)}_s$ transition form factors,
we use the same method in Ref.~\cite{Cheng:2003sm}.
The $D^{(*)}_s$ decay constants are taken to be
%% D^{(*)}_s decay constants
\begin{equation}
  \begin{split}
    f_{D_s} & = 260 \pm 13 \,\text{MeV},\\
    f_{\Dst_s} & = 260 \pm 13 \,\text{MeV}.
  \end{split}
  \label{eq:Ds_decayconst}
\end{equation}
The decay constant of $D_s$ is consistent with the measured
values in Ref.~\cite{PDG}.
The decay constant of $\Dst_s$ should be the same as $D_s$ in heavy-quark limit.
Using these two decay constants as constraints,
we calculate the transition form factor, which is parametrized as
%% FF form
\begin{equation}
  F^{\Bbar_s D^{(*)}_s}(q^2) = \frac{F(0)}{1-aq^2+bq^4}.
  \label{eq:FF_form}
\end{equation}
The three parameters $F(0)$, $a$, and $b$ of different form factors are
given in Table~\ref{table:basic_FF}.

%%%%%%%%%%%%%%%%%%%%%%%%%%%%%%%%%%%%%%%%%%
%% Table IX: Conversion of form factors
\begin{table}[t]
\begin{tabular}{c c c c}
\hline \hline
                  & $F(0)$ & $a$ & $b$ \\
\hline
$F^{\Bbar_s D_s}_0$ & \err{0.67}{0.03} & 0.58 & 0.06 \\
$F^{\Bbar_s D_s}_1$ & \err{0.67}{0.03} & 1.24 & 0.46 \\

\hline
$V^{\Bbar_s \Dst_s}$ & \err{0.77}{0.04} & 1.42 & 0.68 \\
$A^{\Bbar_s \Dst_s}_0$ & \err{0.65}{0.03} & 1.37 & 0.63 \\
$A^{\Bbar_s \Dst_s}_1$ & \err{0.62}{0.03} & 0.77 & 0.11 \\
$A^{\Bbar_s \Dst_s}_2$ & \err{0.59}{0.03} & 1.27 & 0.56 \\
\hline \hline
\end{tabular}
\caption{The transition form factors for $\Bbar_s \to D^{(*)}_s$ used in this work.}
\label{table:basic_FF}
\end{table}
%%%%%%%%%%%%%%%%%%%%%%%%%%%%%%%%%%%%%%%%%%%%%%%%%%

\section{Some Conversion and Transformation of Form Factors}
\label{FF_transform}
Table~\ref{table:FF_transform} provides
the conversion of our notations to
the usual notations of standard form factors.

%%%%%%%%%%%%%%%%%%%%%%%%%%%%%%%%%%%%%%%%%%%%%%%%%%
%% Table X: Conversion of form factors
\begin{table}[ht]
\begin{tabular}{c c c c c}
\hline \hline
                  & $D_s$ & $D_{s0}$ & $\Dst_s$ & $D_{s1}(2460, 2536)$ \\
\hline
$f_{\mathcal{D(*)}s}$ & $f_{Ds}$ & $f_{Ds0}$ & $f_{\Dst s}$ & $-f_{Ds1}$ \\
\hline
$F^{\Bbar_s \mathcal{D}s}_1$ & $F^{\Bbar_s Ds}_1$ & $-F^{\Bbar_s Ds0}_1$ & & \\
$F^{\Bbar_s \mathcal{D}s}_0$ & $F^{\Bbar_s Ds}_0$ & $-F^{\Bbar_s Ds0}_0$ & & \\
\hline
$F^{\Bbar_s \mathcal{\Dst}s}_3$ & & &
  $V^{\Bbar_s \Dst s}$ & $-\frac{m_{Bs}+m_{Ds1}}{m_{Bs}-m_{Ds1}}A^{\Bbar_s \Dst s0}$
\\
$F^{\Bbar_s \mathcal{\Dst}s}_1$ & & &
  $A^{\Bbar_s \Dst s}_1$ & $\frac{m_{Bs}-m_{Ds1}}{m_{Bs}+m_{Ds1}}V^{\Bbar_s \Dst s0}_1$
\\
$F^{\Bbar_s \mathcal{\Dst}s}_2$ & & &
  $A^{\Bbar_s \Dst s}_2$ & $\frac{m_{Bs}+m_{Ds1}}{m_{Bs}-m_{Ds1}}V^{\Bbar_s \Dst s0}_2$
\\
$F^{\Bbar_s \mathcal{\Dst}s}_0$ & & &
  $A^{\Bbar_s \Dst s}_0$ & $V^{\Bbar_s \Dst s0}_0$ \\
\hline \hline
\end{tabular}
\caption{The conversion of the form factors notation in this work to
the usual notation in the literature.
}
\label{table:FF_transform}
\end{table}
%%%%%%%%%%%%%%%%%%%%%%%%%%%%%%%%%%%%%%%%%%%%%%%%%%

If \CP~is conserved,
the form factors of current produced particle pair and antiparticle pair
can be related.
For the standard form factors, the transformation reads
%% eq: Standard FF CP
\begin{equation}
\begin{split}
f_{\Dbar s, \Dbar ^* s, \Dbar s1} = +f_{Ds, D^* s, Ds1}, \quad &
    f_{\Dbar s0, \Dbar s1'}  = -f_{Ds0, Ds1'},\\
F^{Bs\Dbar s}_{0,1}(q^2)  = - F^{\Bbar s Ds}_{0,1}(q^2), \quad &
    F^{Bs\Dbar s0}_{0,1}(q^2)  = + F^{\Bbar s Ds0}_{0,1}(q^2),\\
F^{Bs(\Dbar ^* s, \Dbar s1)}_{0,1,2}(q^2)
        = - F^{\Bbar s (D^*s, Ds1)}_{0,1,2}(q^2), \quad &
F^{Bs\Dbar s1'}_{0,1,2}(q^2)
        = + F^{\Bbar s Ds1'}_{0,1,2}(q^2), \\
F^{Bs(\Dbar ^* s, \Dbar s1)}_{3}(q^2)
        = + F^{\Bbar s (D^*s, Ds1)}_{3}(q^2), \quad &
F^{Bs\Dbar s1'}_{3}(q^2)
        = - F^{\Bbar s Ds1'}_{3}(q^2), \\
\end{split}
\label{eq:Std_CP}
\end{equation}
where $D_{s1}$ and $D_{s1'}$ are the $\CP$-even and $\CP$-odd states
of the linear combination of $D_{s1}(2460)$ and $D_{s1}(2536)$.
The relations for form factors in Eq.~\eqref{eq:PP_Cur_Form} to
Eq.~\eqref{eq:VV_Cur_Form} are
%% eq: Time-like CP
\begin{equation}
\begin{split}
F^{PP}_{0,1}(q^2) & = - F^{\overline{PP}}_{0,1}(q^2), \\
V^{VP,VV}(q^2) & = + V^{\overline{VP}, \overline{VV}}(q^2), \\
A^{VP,VV}(q^2) & = - A^{\overline{VP}, \overline{VV}}(q^2).
\end{split}
\label{eq:Cur_CP}
\end{equation}
The transformations for transition form factors from Eq.~\eqref{eq:PP_Tr_Form}
to Eq.~\eqref{eq:VV_Tr_Form} are
%% eq: Transition CP
\begin{equation}
\begin{split}
V^{\Bbar_s PP,\Bbar_s VP,\Bbar_s VV}  &
        = - V^{B_s\overline{PP}, B_s\overline{PP}, B_s\overline{PP}},  \\
A^{\Bbar_s PP,\Bbar_s VP,\Bbar_s VV}  &
        = + A^{B_s\overline{PP}, B_s\overline{VP}, B_s\overline{VV}}.
\end{split}
\label{eq:Tr_CP}
\end{equation}
Compared with Eq.~(\ref{eq:Cur_CP}), there is
one additional minus sign coming from the pseudoscalar $B_s$ meson.

\section{Pole Contribution to Form Factors}
\label{full_pole}

For simplicity, we only list the contributions from $D_s$ and $\Dst_s$ poles.
The contributions of $D_{sJ}(2700)$ have the same forms as $\Dst_s$,
but with different mass, width, and strong coupling constants.

In the time-like $DK$ transition form factors,
$D^*_s$ is the only possible pole.
But there is an ambiguity in the matrix element
$\langle DK |i \mathcal{L}_{\rm eff} |D^{*}_{\rm int}\rangle$ when $D^{*}$ goes to
offshell. The matrix element is given by
%% Total Form Factors
\begin{equation}
\langle D(p_D)K(p_K) |i \mathcal{L}_{\rm eff} |D^{*}_{\rm int}(p_{D^*},\eps_{D^{*}})\rangle
= \eps_{\rm int} \cdot (\frac{1}{2}(p_K-p_D)+\alpha q),
 \label{DKDst_matrix}
\end{equation}
where $\alpha$ is undetermined since
the associated term is zero when $D^*_s$ is on-shell.
According to this matrix element,
the pole contribution to time-like form factor becomes
%% DK pole time like
\begin{equation}
\begin{split}
F_1^{DK}(q^2) = &
  \frac{g_{\Dst DP}f_{\Dst_{\rm int}}m_{\rm int*}}{q^2-m^2_{\rm int*}+im_{\rm int*}\Gamma_{\rm int*}}
  \frac{1}{2}, \\
F_0^{DK}(q^2) = &
  \frac{g_{\Dst DP}f_{\Dst_{\rm int}}m_{\rm int*}}{q^2-m^2_{\rm int*}+im_{\rm int*}\Gamma_{\rm int*}}
  (\frac{q^2-m^2_{\rm int*}}{m_{\rm int*}^2}
  (\frac{q^2}{m_{D}^2-m_{K}^2}\alpha- \frac{1}{2})),\\
\end{split}
 \label{DK_Cur_pole}
\end{equation}
where $m_{\rm int*}$ and $\Gamma_{\rm int*}$ are the mass and width of the $\Dst_s$ pole,
respectively. If $\alpha$ is nonzero, $A_0^{DK}(q^2)$ will increase as $q^2$ increase.
Such energy dependence is unnatural for form factors.
We hence set $\alpha$ as zero. Once $\alpha$ is fixed, we have the following
pole contribution to transition form factors
%% DK pole transition
\be
 \label{DK_Tr_pole}
    \frac{V^{\Bbar_s DK} }{m_{B_s}^3}
&= &
    (\frac{g_{\Dst DP}}{q^2-m^2_{\rm int*}+im_{\rm int*}\Gamma_{\rm int*}})
    \frac{1}{2}\frac{2V^{\Bbar_s \Dst}}{m_{B_s}+m_{\rm int*}},
\non\\
    \frac{A_1^{\Bbar_s DK} }{m_{B_s}}
&=&
    (\frac{g_{\Dst DP}}{q^2-m^2_{\rm int*}+im_{\rm int*}\Gamma_{\rm int*}})
    \frac{q'^2}{2(q'^2+q^2-m_{B_s}^2)}(q'(p_D-p_K)-\frac{m_D^2-m_K^2}{m_{\rm int*}^2}qq') \times
\non\\
&&   \quad (\frac{m_{B_s}+m_{\rm int*}}{q'^2}A^{\Bbar_s \Dst}_1+
     (1-\frac{m_{B_s}^2-m_{\rm int*}^2}{q'^2}) \frac{A^{\Bbar_s \Dst}_2}{m_{B_s}+m_{\rm int*}}-
     \frac{2m_{\rm int*}}{q'^2}A^{\Bbar_s \Dst}_0),
\\
    \frac{A_2^{\Bbar_s DK} }{m_{B_s}}
&=&
    (\frac{g_{\Dst DP}}{q^2-m^2_{\rm int*}+im_{\rm int*}\Gamma_{\rm int*}})
    \frac{-1}{2}(m_{B_s}+m_{\rm int*})A^{\Bbar_s \Dst}_1,
\non \\
    \frac{A_0^{\Bbar_s DK} }{m_{B_s}}
&=&
    (\frac{g_{\Dst DP}}{q^2-m^2_{\rm int*}+im_{\rm int*}\Gamma_{\rm int*}})
    \lbrace \frac{q^2}{m^2_D-m^2_K}[
      \frac{m_D^2-m_K^2}{2m_{\rm int*}^2}(m_{B_s}+m_{\rm int*})A^{\Bbar_s \Dst}_1
\non\\
&&     +(q'(p_D-p_K)-\frac{m_D^2-m_K^2}{m_{\rm int*}^2}qq')
      \frac{A^{\Bbar_s \Dst}_2}{m_{B_s}+m_{\rm int*}}
      -2\frac{A_1^{\Bbar_s DK}}{m_{B_s}}] +
    \frac{A_2^{\Bbar_s DK}}{m_{B_s}} \rbrace,
\non
\en
where $q=p_{D}+p_{K}$ is the total momentum of transitioned mesons,
and $q'=p_{\Bbar_s}-q$ is the momentum of weak current.

Other modes receive contribution from both $D_s$ and $\Dst_s$ poles.
The time-like form factors of $\Dst K$ are
%% DstK pole current
\be
 \label{DstK_Cur_pole}
  \frac{2V^{\Dst K}(q^2)}{m_{\Dst}+m_K}
&= &
    (\frac{-g_{\Dst \Dst P}f_{\Dst_{\rm int}}m_{\rm int*}}{q^2-m^2_{\rm int*}+im_{\rm int*}\Gamma_{\rm int*}}),
\non\\
    A_1^{\Dst K}(q^2)
&= & 0,
\non\\
   A_2^{\Dst K}(q^2)
&= & 0,
\\
   2 m_{\Dst} A_0^{\Dst K}(q^2)
&= &
    (\frac{g_{\Dst DP}f_{D_{\rm int}}}{q^2-m^2_{\rm int}+im_{\rm int}\Gamma_{\rm int}})q^2,
\non
\en
where $m_{\rm int(*)}$ and $\Gamma_{\rm int(*)}$ are the mass and width of
the $D^{(*)}_s$ poles respectively. The $\Bbar_s$ to $\Dst K$
transition form factors induced by $D^{(*)}_s$ poles are given by
%% DstK pole transition
\be
 \label{DstK_Tr_pole}
\frac{V^{\Bbar_s \Dst K}_{2} }{m_{B_s}^2}
& = &
    (\frac{g_{\Dst \Dst P}}
     {q^2-m^2_{\rm int*}+im_{\rm int*}\Gamma_{\rm int*}})
    (m_{B_s}+m_{\rm int*})A_1^{\Bbar_s \Dst},
\non \\
\frac{V^{\Bbar_s \Dst K}_{1} }{m_{B_s}^4}
&= &
    (\frac{g_{\Dst \Dst P}}
     {q^2-m^2_{\rm int*}+im_{\rm int*}\Gamma_{\rm int*}})
    \frac{A_2^{\Bbar_s \Dst}}{(m_{B_s}+m_{\rm int*})},
\non\\
\frac{V^{\Bbar_s \Dst K}_{0} }{m_{B_s}^2}
& = &
    (\frac{-g_{\Dst \Dst P}}
     {q^2-m^2_{\rm int*}+im_{\rm int*}\Gamma_{\rm int*}})
    (2m_{\rm int*})A_0^{\Bbar_s \Dst}
\non\\
&&   -(q^2-m_{\rm int*}^2)\frac{V^{\Bbar_s \Dst K}_{1} }{m_{B_s}^4},
\\
\frac{A^{\Bbar_s \Dst K}_{3} }{m_{B_s}^4}
& = &
    (\frac{g_{\Dst \Dst P}}
     {q^2-m^2_{\rm int*}+im_{\rm int*}\Gamma_{\rm int*}})
    \frac{2V^{\Bbar_s \Dst}}{(m_{B_s}+m_{\rm int*})},
\non\\
\frac{A_1^{\Bbar_s \Dst K} }{m_{B_s}^2}
&= &
    (\frac{-g_{\Dst D P}}
     {q^2-m^2_{\rm int}+im_{\rm int}\Gamma_{\rm int}})
    A_1^{\Bbar_s D},
\non\\
\frac{A_0^{\Bbar_s \Dst K} }{m_{B_s}^2}
&= &
    (\frac{m_{B_s}^2-m_{\rm int}^2}{m_{B_s}^2-q^2})
    (\frac{-g_{\Dst DP}}
     {q^2-m^2_{\rm int}+im_{\rm int}\Gamma_{\rm int}})
    A_0^{\Bbar_s D}
\non\\
&&   -(\frac{q^2-m_{\rm int}^2}{m_{B_s}^2-q^2})
    \frac{A_1^{\Bbar_s \Dst K} }{m_{B_s}^2}.
\non
\en
The $D\Kst$ and $\Dst K$ form factors are parameterized in the same way.
The pole part of the $D\Kst$ time-like form factors are
%% DKst pole current
\be
\label{DKst_Cur_pole}
\frac{2V^{D\Kst}(q^2)}{m_{D}+m_{\Kst}} &= &
    (\frac{4f_{\Dst DV}f_{\Dst_{\rm int}}m_{\rm int*}}{q^2-m^2_{\rm int*}+im_{\rm int*}\Gamma_{\rm int*}}),
\non\\
A_1^{D \Kst}(q^2) &= & 0, \non\\
A_2^{D \Kst}(q^2) &= & 0, \\
2 m_{\Kst} A_0^{D \Kst}(q^2) &= &
    -2q^2(\frac{g_{DDV}f_{D_{\rm int}}}{q^2-m^2_{\rm int}+im_{\rm int}\Gamma_{\rm int}}). \non
\en
And the transition form factors derived from the pole model are written as
%% DKst pole transition
\be
 \label{DKst_Tr_pole}
\frac{V^{\Bbar_s D\Kst}_{2} }{m_{B_s}^2} &= &
    (\frac{-4f_{\Dst DV}}
     {q^2-m^2_{\rm int*}+im_{\rm int*}\Gamma_{\rm int*}})
    (m_{B_s}+m_{\rm int*})A_1^{\Bbar_s \Dst},
\non    \\
\frac{V^{\Bbar_s D\Kst}_{1} }{m_{B_s}^4}
&= &
    (\frac{-4f_{\Dst DV}}
     {q^2-m^2_{\rm int*}+im_{\rm int*}\Gamma_{\rm int*}})
    \frac{A_2^{\Bbar_s \Dst}}{(m_{B_s}+m_{\rm int*})},
\non   \\
\frac{V^{\Bbar_s D\Kst}_{0} }{m_{B_s}^2}
&= &
    (\frac{4f_{\Dst DV}}
     {q^2-m^2_{\rm int*}+im_{\rm int*}\Gamma_{\rm int*}})
    (2m_{\rm int*})A_0^{\Bbar_s \Dst}
\non    \\
&&   -(q^2-m_{\rm int*}^2)\frac{V^{\Bbar_s D\Kst}_{1} }{m_{B_s}^4},
\\
\frac{A^{\Bbar_s D\Kst}_{3} }{m_{B_s}^4}
&= &
    (\frac{-4f_{\Dst DV}}
     {q^2-m^2_{\rm int*}+im_{\rm int*}\Gamma_{\rm int*}})
    \frac{2V^{\Bbar_s \Dst}}{(m_{B_s}+m_{\rm int*})},
\non    \\
\frac{A_1^{\Bbar_s D\Kst} }{m_{B_s}^2}
&= &
    (\frac{2g_{DDV}}
     {q^2-m^2_{\rm int}+im_{\rm int}\Gamma_{\rm int}})
    A_1^{\Bbar_s D},
\non    \\
\frac{A_0^{\Bbar_s D\Kst} }{m_{B_s}^2}
& = &
    (\frac{m_{B_s}^2-m_{\rm int}^2}{m_{B_s}^2-q^2})
    (\frac{2g_{DDV}}
     {q^2-m^2_{\rm int}+im_{\rm int}\Gamma_{\rm int}})
    A_0^{\Bbar_s D}
\non    \\
&&   -(\frac{q^2-m_{\rm int}^2}{m_{B_s}^2-q^2})
    \frac{A_1^{\Bbar_s D\Kst} }{m_{B_s}^2}.
\non
\en

Finally, the $\Dst \Kst$ time-like form factors
from $D^{(*)}_s$ poles are
%% DstKst pole current
\be
 \label{DstKst_Cur_pole}
\frac{V^{\Dst \Kst}_0(q^2)}{(m_{\Dst}+m_{\Kst})^2}
& = &
    (\frac{-4f_{\Dst DV} f_{\rm int}}
     {q^2-m^2_{\rm int}+im_{\rm int}\Gamma_{\rm int}}),
 \non\\
\frac{V^{\Dst \Kst}_1(q^2)}{(m_{\Dst}+m_{\Kst})^2}
&= & 0,
\non\\
\frac{V^{\Dst \Kst}_2(q^2)}{(m_{\Dst}+m_{\Kst})^2}
&= & 0,
\non\\
A_{11}^{\Dst \Kst}(q^2)
&= &
    (\frac{2g_{\Dst \Dst V} m_{\rm int*}f_{\rm int*}}
     {q^2-m^2_{\rm int*}+im_{\rm int*}\Gamma_{\rm int*}}),
\non     \\
A_{12}^{\Dst \Kst}(q^2)
&= &
    (\frac{-4f_{\Dst \Dst V} m_{\rm int*}f_{\rm int*}}
     {q^2-m^2_{\rm int*}+im_{\rm int*}\Gamma_{\rm int*}}),
\non     \\
A_2^{\Dst \Kst}(q^2)
&= &
    (\frac{-2f_{\Dst \Dst V} m_{\rm int*}f_{\rm int*}}
     {q^2-m^2_{\rm int*}+im_{\rm int*}\Gamma_{\rm int*}}),
\\
A_{01}^{\Dst \Kst}(q^2)
&= &
    (\frac{-2g_{\Dst \Dst V} m_{\rm int*}f_{\rm int*}}
     {q^2-m^2_{\rm int*}+im_{\rm int*}\Gamma_{\rm int*}})
     \frac{q^2}{m_{\rm int*}^2}
     +(\frac{4f_{\Dst \Dst V} m_{\rm int*}f_{\rm int*}}
     {q^2-m^2_{\rm int*}+im_{\rm int*}\Gamma_{\rm int*}})
     \frac{q^2}{m_{\rm int}^2}
\non    \\
&&   +A_{11}^{\Dst \Kst}(q^2)+A_{12}^{\Dst \Kst}(q^2),
\non\\
A_{02}^{\Dst \Kst}(q^2)
&= &
    (\frac{4f_{\Dst \Dst V} m_{\rm int*}f_{\rm int*}}
     {q^2-m^2_{\rm int*}+im_{\rm int*}\Gamma_{\rm int*}})
     (\frac{1}{2}-\frac{q^2-m_{\Dst}^2+m_{\Kst}^2}{2m_{\rm int*}^2})
     \frac{q^2 }{(m_{\Dst}+m_{\Kst})^2}
\non     \\
&&   +(\frac{m_{\Dst}^2-m_{\Kst}^2}{q^2})A_2^{\Dst \Kst}(q^2).
\non
\en
And the transition form factors are given by the following three equations.
The first part is the form factors from vector current
%% DstKst pole transition: V
\be
 \label{DstKst_Tr_pole_V}
\frac{V^{\Bbar_s \Dst \Kst}_{3} }{m_{B_s}^3} &= &
    (\frac{4f_{\Dst \Dst V}}
     {q^2-m^2_{\rm int*}+im_{\rm int*}\Gamma_{\rm int*}})
     \frac{2V^{\Bbar_s \Dst}}{m_{B_s}+m_{\rm int*}},
     \non\\
\frac{V^{\Bbar_s \Dst \Kst}_{2} }{m_{B_s}^3} &= &
    (\frac{2g_{\Dst \Dst V}}
     {q^2-m^2_{\rm int*}+im_{\rm int*}\Gamma_{\rm int*}})
     \frac{2V^{\Bbar_s \Dst}}{m_{B_s}+m_{\rm int*}},
     \non\\
\frac{V^{\Bbar_s \Dst \Kst}_{1} }{m_{B_s}^3} &= &
    (\frac{4f_{\Dst \Dst V}}
     {q^2-m^2_{\rm int*}+im_{\rm int*}\Gamma_{\rm int*}})
     (-\frac{2V^{\Bbar_s \Dst}}{m_{B_s}+m_{\rm int*}}),
     \\
\frac{V^{\Bbar_s \Dst \Kst}_{01} }{m_{B_s}^3} &= &
    (\frac{-4f_{\Dst DV}}
     {q^2-m^2_{\rm int}+im_{\rm int}\Gamma_{\rm int}})
     A^{\Bbar_s D}_1,
     \non\\
\frac{V^{\Bbar_s \Dst \Kst}_{00} }{m_{B_s}^3} &= &
    (\frac{-4f_{\Dst DV}}
     {q^2-m^2_{\rm int}+im_{\rm int}\Gamma_{\rm int}})
     A^{\Bbar_s D}_0.
     \non
\en
The second part which originate from axial currents are
%% DstKst pole transition: A1
\be
\label{DstKst_Tr_pole_A1}
\frac{A_{62}^{\Bbar_s \Dst \Kst} }{m_{B_s}}
&= &
    (\frac{4f_{\Dst \Dst V}}
     {q^2-m^2_{\rm int*}+im_{\rm int*}\Gamma_{\rm int*}})
     \frac{1}{2}(m_{B_s}+m_{\rm int*})A^{\Bbar_s \Dst}_{1},
\non     \\
\frac{A_{61}^{\Bbar_s \Dst \Kst} }{m_{B_s}}
&= &
    (\frac{4f_{\Dst \Dst V}}
     {q^2-m^2_{\rm int*}+im_{\rm int*}\Gamma_{\rm int*}}) \times
\non     \\
&&    \quad \lbrace
      -(-q'p_{\Kst}+\frac{qq' \cdot qp_{\Kst}}{m_{\rm int*}^2})
       \frac{A^{\Bbar_s \Dst}_2}{m_{B_s}+m_{\rm int*}}+
     \frac{qp_{\Kst}}{2m_{\rm int*}^2}(m_{B_s}+m_{\rm int*})A^{\Bbar_s \Dst}_{1}
     \rbrace
\non     \\
&&    -\frac{1}{2}(1-\frac{m_{\Dst}^2-m_{\Kst}^2}{q^2})
     \frac{A_{62}^{\Bbar_s \Dst \Kst} }{m_{B_s}},
\non     \\
%\en
%%%%%%%%%%%%%%%%%%%%%%%%%%%%%%%%%
%\be
%\label{DstKst_Tr_pole_A2}
m_{B_s} A_{60}^{\Bbar_s \Dst \Kst}
&= &
    (\frac{4f_{\Dst \Dst V}}
     {q^2-m^2_{\rm int*}+im_{\rm int*}\Gamma_{\rm int*}})
     (-q'p_{\Kst}+\frac{qq' \cdot qp_{\Kst}}{m_{\rm int*}^2}) \times
     \\
&&    \quad \lbrace
       -(m_{B_s}+m_{\rm int*})A^{\Bbar_s \Dst}_{1}+2m_{\rm int*}A^{\Bbar_s \Dst}_{0}
       -(q'^2-(m_{B_s}^2-m_{\rm int*}^2))\frac{A^{\Bbar_s \Dst}_2}{m_{B_s}+m_{\rm int*}}
     \rbrace,
\non     \\
&&   -(q'^2-(m_{B_s}^2-q^2))\frac{A_{61}^{\Bbar_s \Dst \Kst} }{m_{B_s}},
\non \\
\frac{A_{3}^{\Bbar_s \Dst \Kst} }{m_{B_s}}
&= &
    (\frac{-2g_{\Dst \Dst V} m_{\rm int*}f_{\rm int*}}
     {q^2-m^2_{\rm int*}+im_{\rm int*}\Gamma_{\rm int*}})
     (m_{B_s}+m_{\rm int*})A^{\Bbar_s \Dst}_{1},
\non     \\
\frac{A_{4}^{\Bbar_s \Dst \Kst} }{m_{B_s}}
&= &
    (\frac{4f_{\Dst \Dst V}}
     {q^2-m^2_{\rm int*}+im_{\rm int*}\Gamma_{\rm int*}})
     (m_{B_s}+m_{\rm int*})A^{\Bbar_s \Dst}_{1},
\non     \\
\frac{A_{21}^{\Bbar_s \Dst \Kst} }{m_{B_s}^3}
& = &
    (\frac{-2g_{\Dst \Dst V} m_{\rm int*}f_{\rm int*}}
     {q^2-m^2_{\rm int*}+im_{\rm int*}\Gamma_{\rm int*}})
     \frac{-A^{\Bbar_s \Dst}_2}{m_{B_s}+m_{\rm int*}},
\non     \\
\frac{A_{20}^{\Bbar_s \Dst \Kst} }{m_{B_s}}
& = &
    (\frac{-2g_{\Dst \Dst V} m_{\rm int*}f_{\rm int*}}
     {q^2-m^2_{\rm int*}+im_{\rm int*}\Gamma_{\rm int*}}) \times
\non     \\
& &   \quad \lbrace
       -(m_{B_s}+m_{\rm int*})A^{\Bbar_s \Dst}_{1}
       -(q'^2-(m_{B_s}^2-m_{\rm int*}^2))\frac{A^{\Bbar_s \Dst}_2}{m_{B_s}+m_{\rm int*}}
       +2m_{\rm int*}A^{\Bbar_s \Dst}_{0}
     \rbrace
\non     \\
&&   -(q'^2-(m_{B_s}^2-q^2))\frac{A_{21}^{\Bbar_s \Dst \Kst} }{m_{B_s}^3},
\non
\en
and
%% DstKst pole transition: A3
\be
\label{DstKst_Tr_pole_A3}
\frac{A_{11}^{\Bbar_s \Dst \Kst} }{m_{B_s}^3}
&= &
    (\frac{4f_{\Dst \Dst V}}
     {q^2-m^2_{\rm int*}+im_{\rm int*}\Gamma_{\rm int*}})
     \frac{-A^{\Bbar_s \Dst}_2}{m_{B_s}+m_{\rm int*}},
\non     \\
\frac{A_{10}^{\Bbar_s \Dst \Kst} }{m_{B_s}}
&= &
    (\frac{4f_{\Dst \Dst V}}
     {q^2-m^2_{\rm int*}+im_{\rm int*}\Gamma_{\rm int*}})
\non    \\
&&    \quad \lbrace
       -(m_{B_s}+m_{\rm int*})A^{\Bbar_s \Dst}_{1}
       -(q'^2-(m_{B_s}^2-m_{\rm int*}^2))\frac{A^{\Bbar_s \Dst}_2}{m_{B_s}+m_{\rm int*}}
       +2m_{\rm int*}A^{\Bbar_s \Dst}_{0}
     \rbrace
\\
&&   -(q'^2-(m_{B_s}^2-q^2))\frac{A_{11}^{\Bbar_s \Dst \Kst} }{m_{B_s}^3},
\non\\
\frac{A_{01}^{\Bbar_s \Dst \Kst} }{m_{B_s}^3}
&= &
    \lbrace
     \frac{-2g_{\Dst \Dst V} m_{\rm int*}f_{\rm int*}}
     {q^2-m^2_{\rm int*}+im_{\rm int*}\Gamma_{\rm int*}}
     +\frac{4f_{\Dst \Dst V}}
     {q^2-m^2_{\rm int*}+im_{\rm int*}\Gamma_{\rm int*}}
    \rbrace \times
\non    \\
&&   \quad \lbrace
     \frac{-1}{2m_{\rm int*}^2}(m_{B_s}+m_{\rm int*})A^{\Bbar_s \Dst}_{1}
     +\frac{qq'}{m_{\rm int*}^2}\frac{A^{\Bbar_s \Dst}_2}{m_{B_s}+m_{\rm int*}}
    \rbrace
\non    \\
&&   +\frac{1}{2q^2}(\frac{A_{3}^{\Bbar_s \Dst \Kst} }{m_{B_s}}+
     \frac{A_{4}^{\Bbar_s \Dst \Kst} }{m_{B_s}}),
\non     \\
\frac{A_{00}^{\Bbar_s \Dst \Kst} }{m_{B_s}}
&= &
    \lbrace
     \frac{-2g_{\Dst \Dst V} m_{\rm int*}f_{\rm int*}}
     {q^2-m^2_{\rm int*}+im_{\rm int*}\Gamma_{\rm int*}}
     +\frac{4f_{\Dst \Dst V}}
     {q^2-m^2_{\rm int*}+im_{\rm int*}\Gamma_{\rm int*}}
    \rbrace \times
\non    \\
&&   \quad \frac{qq'}{m_{\rm int*}^2}
     \lbrace
     (m_{B_s}+m_{\rm int*})A^{\Bbar_s \Dst}_{1}
      +(q'^2-(m_{B_s}^2-m_{\rm int*}^2))\frac{A^{\Bbar_s \Dst}_2}{m_{B_s}+m_{\rm int*}}
      -2m_{\rm int*}A^{\Bbar_s \Dst}_{0}
     \rbrace
\non     \\
&&   -(q'^2-(m_{B_s}^2-q^2))\frac{A_{01}^{\Bbar_s \Dst \Kst} }{m_{B_s}^3}.
\non
\en

%%%%%%%%%%%%%%%%%%%%%%%%%%%%%%%%%%%%%%%%%%%%%%%%%%%%%%%

%%%%%%%%%%%%%%%%%%%%%%%%%%%

%%
\end{document}